\documentclass[prd, preprint, nofootinbib]{revtex4}%

\usepackage{amsfonts}
\usepackage{amsmath}
\usepackage{amssymb}
\usepackage{graphicx}
\usepackage{appendix}
\usepackage{color}

\newcommand{\dd}{{\rm{d}}} 

\begin{document}


\begin{flushright}
$LIFT$--8-1.25
\end{flushright}
\vspace{-6.5mm}


\title{\LARGE
Revisiting black holes of algebraic type D with a cosmological constant\\[5mm]}

\author{Hryhorii Ovcharenko}
\email{hryhorii.ovcharenko@matfyz.cuni.cz}
\affiliation{Charles University, Faculty of Mathematics and Physics,
Institute of Theoretical Physics,
V~Hole\v{s}ovi\v{c}k\'ach 2, 18000 Prague 8, Czechia}

\author{Ji\v{r}\'i Podolsk\'{y}}
\email{jiri.podolsky@mff.cuni.cz}  
\affiliation{Charles University, Faculty of Mathematics and Physics,
Institute of Theoretical Physics,
V~Hole\v{s}ovi\v{c}k\'ach 2, 18000 Prague 8, Czechia}

\author{Marco Astorino}
\email{marco.astorino@gmail.com}
\affiliation{Laboratorio Italiano di Fisica Teorica (LIFT), \\
Via Archimede 20, I-20129 Milano, Italy}

\begin{abstract}
As an extension of our previous work \cite{OvcharenkoPodolskyAstorino:2024}, we study a complete family of type D black holes with Kerr-like rotation, NUT twist, acceleration, electric and magnetic charges, and any value of the cosmological constant~$\Lambda$. We relate various metric forms of these spacetimes, namely those found by Pleba\'nski-Demia\'nski (PD), Griffiths-Podolsk\'y (GP), and most recently Astorino (A). By explicit coordinate transformations and proper  identification of the physical parameters we show that these representations are locally equivalent, and cover the entire class of type D solutions of the Einstein-Maxwell-$\Lambda$ equations, such that the (non-null) electromagnetic field is aligned with both the (double-degenerate) principal null directions of the Weyl tensor. In particular, we concentrate on the subclass which describes accelerating NUT black holes without the Kerr-like rotation.
\end{abstract}

\date{\today}


\maketitle

\tableofcontents
\newpage

\section{Introduction}

Exact solutions to the field equations of general relativity representing black holes were found soon after Einstein formulated his theory of gravity --- although it took several decades to fully understand their geometrical properties, global structure and physical meaning. The first of them was spherical vacuum solution by Schwarzschild (1916), followed by the Reissner-Nordstr\"{o}m solution with an electric charge (1916, 1918), and the Kottler-Weyl-Trefftz solution with a cosmological constant~$\Lambda$ (1918--1922). Half a century later these were extended by the discovery of rotating Kerr (1963), twisting Taub-NUT (Newman, Tamburino, Unti, 1963) and Kerr-Newman charged black holes (1965). Also acceleration was added to them after Kinnersley and Walker (1970) properly interpreted the previously known $C$-metric (1918, 1962).

All these famous black holes are spherically/axially symmetric exact solutions of the field equations. Even more interestingly, they belong to a general family of algebraic type~D spacetimes with any cosmological constant~${\Lambda}$ and (double-aligned, non-null) electromagnetic field. Such a family was found in 1971 by Debever \cite{Debever:1971}, but its more convenient metric form was presented in 1976 by Pleba\'nski and Demia\'nski \cite{PlebanskiDemianski:1976} --- review and specific references can be found, e.g., in the monographs \cite{Stephanietal:2003} and \cite{GriffithsPodolsky:2009}, or in \cite{VandenBergh:2017}.

At about 2005, the large Pleba\'nski-Demia\'nski (PD) class of spacetimes was investigated in more detail by
Griffiths and Podolsk\'y (GP) \cite{GriffithsPodolsky:2005, GriffithsPodolsky:2006, PodolskyGriffiths:2006}, and a few years ago again by Podolsk\'y and Vr\'atn\'y (PV) \cite{PodolskyVratny:2021, PodolskyVratny:2023}. Recently, these type~D spacetimes were reconsidered by Astorino~(A) using solution generating techniques, and a novel metric form was thus obtained \cite{Astorino:2024b, Astorino:2024a}. Its main advantage is that it directly admits limits to \emph{all the subcases}, including the peculiar accelerating solutions with (just) the NUT parameter, which  was previously considered to exist outside the type~D class \cite{ChngMannStelea:2006,PodolskyVratny:2020}, see also \cite{Astorino:2023elf, Astorino:2023b}.  Also, after the presentation of the new A metric it was not obvious whether it is fully equivalent to the PD metric.

This was the motivation of our work \cite{OvcharenkoPodolskyAstorino:2024}. In it, we put the A metric into a more compact A$^+$ form, and then elucidated its relations to PD, GP, and PV metric representations of this big family of black hole solutions in the Einstein-Maxwell-$\Lambda$ theory. However, due to considerable complexity of the exact relation between the A$^+$ and PD forms, in \cite{OvcharenkoPodolskyAstorino:2024} we restricted the analysis only to the case when the ${\Lambda=0}$. In the present contribution we extend such a study to the most general situation in which any value of the cosmological constant $\Lambda$ is allowed.

We will use the same nomenclature and conventions for A$^+$, PD, GP, and PV coordinates and parameters as in our previous paper \cite{OvcharenkoPodolskyAstorino:2024} which concentrated on black holes with ${\Lambda=0}$. These are summarized in Table~\ref{Tab-summary-of-metrics}.

\vspace{6mm}
\begin{table}[!h]
\begin{center}

\begin{tabular}{l|c|c|c|c}
\hline
\hline
Metric form\  & \ Equation\  & \ Source\  &  \ Coordinates\ & \  Parameters\ \\[2pt]
\hline
\hline
A$^+$             & \eqref{ds2_simpl}    & \cite{OvcharenkoPodolskyAstorino:2024}
   & ${(r, x, t, \varphi)}$ &   ${\alpha, a, l, m, e, g, \Lambda }$\\
\hline
PD$_{\alpha}$ & \eqref{PD-GP-form}  & \cite{GriffithsPodolsky:2005, GriffithsPodolsky:2006}
   & ${(r', x', \tau', \phi')}$ &   ${\alpha', k', n', \epsilon', m', e', g', \Lambda }$ \\
\hline
GP     & \eqref{GP-metric} & \cite{GriffithsPodolsky:2005, GriffithsPodolsky:2006, PodolskyGriffiths:2006}
   & ${(\tilde{r}, \theta, t, \varphi)}$ & ${\tilde{\alpha}, \tilde{a}, \tilde{l}, \tilde{m}, \tilde{e}, \tilde{g}, \Lambda }$ \\
       & \eqref{GP-metric-final} &
   & ${(\bar{r}, \theta, \bar{t}, \bar{\varphi})}$ & ${\bar{\alpha}, \bar{a}, \bar{l}, \bar{m}, \bar{e}, \bar{g}, \Lambda }$ \\
\hline
\hline
\end{tabular}
\vspace{4.0mm}
\caption{\label{Tab-summary-of-metrics} Summary of the metrics, coordinates and parameters for the most general black-hole spacetimes of algebraic type~D studied in this paper. This is a simplified version of Table I of \cite{OvcharenkoPodolskyAstorino:2024} in which the ${\Lambda=0}$ black holes were written in the A, A$^+$, PD, PD$_{\alpha}$, PD$_{\alpha\omega}$, GP$_{\omega}$, PV$_{\omega}$ metric forms, and their mutual relations presented.}
\end{center}
\end{table}

\vspace{4mm}

In Section~\ref{sc:PD-form} we start with the compact A$^+$ form of the complete Astorino solution \cite{Astorino:2024b}, as was derived in \cite{OvcharenkoPodolskyAstorino:2024}. Then we present the explicit coordinate transformation to the Pleba\'nski-Demia\'nski metric PD$_{\alpha}$, together with relations of the physical Astorino parameters to the PD integration constants. We also discuss various special cases. In subsequent Section~\ref{sc:GP-form} we give the transformation to the Griffiths-Podolsk\'y form GP of this family of spacetimes, again discussing the special cases. Summary and concluding remarks are given in Section~\ref{sc:conclusions}.

\newpage

\section{Transformation from the Astorino to the Pleba\'nski-Demia\'nski metric}
\label{sc:PD-form}

In our previous paper \cite{OvcharenkoPodolskyAstorino:2024} we put the novel Astorino metric \cite{Astorino:2024b}, representing the most general type~D black hole with any cosmological constant $\Lambda$ and doubly-aligned, non-null electromagnetic field, to a more compact form. Such a metric, which we denoted as~A$^+$, reads
\begin{equation}
    \dd s^2=\dfrac{1}{\Omega^2}\bigg[-\dfrac{\Delta_r}{\rho^2}(A\,\dd t - B\,\dd\varphi)^2
    + \dfrac{\Delta_x}{\rho^2}(C\,\dd t + D\,\dd\varphi)^2
    + C_f\,\rho^2 \Big(\,\dfrac{\dd r^2}{\Delta_r} + \dfrac{\dd x^2}{\Delta_x}\,\Big)\bigg],
    \label{ds2_simpl}
\end{equation}
where the functions are
\begin{align}
    \Omega(r,x) &= 1-\alpha\, r\, x \,,\label{Om_cf}\\
    A(x) &= 1 + \alpha^2(l^2-a^2)\,x^2,\label{Aw}\\
    B(x) &= a + 2l\,x + a\,x^2,\label{Bw}\\
    C(r) &= a + 2\alpha l\,r + \alpha^2 a\, r^2,\label{Cw}\\
    D(r) &= (l^2-a^2) + r^2,\label{Bu}
\end{align}
and
\begin{align}
\rho^2 =&\  AD+BC \nonumber\\
       =&\  (l+a\,x)^2 + 2\alpha\,l\,(a+2l\,x+a\,x^2)\,r + \alpha^2(a^2-l^2)^2\,x^2 + [1+\alpha^2(a+l\,x)^2\,]\,r^2 ,\label{rho2expl}\\
\Delta_r=&\ (1-\alpha^2r^2)\big[(r-m)^2 - (m^2+l^2-a^2-e^2-g^2)\big] \nonumber\\
   & \hspace{20mm} - \frac{\Lambda}{3}\Big(\,\frac{3l^2}{1+\alpha^2 a^2}\,r^2
     + \frac{4\alpha a l}{1+\alpha^2 a^2}\,r^3 + r^4 \Big) ,\label{delta_r_init}\\[1mm]
\Delta_x=&\ (1-x^2)\big[(1-\alpha m \,x)^2-\alpha^2x^2(m^2+l^2-a^2-e^2-g^2)\big] \nonumber\\
   & \hspace{20mm} - \frac{\Lambda}{3}\Big(\,\frac{3l^2}{1+\alpha^2 a^2}\,x^2
    +\frac{4al}{1+\alpha^2 a^2}\,x^3 + \frac{a^2+\alpha^2(a^2-l^2)^2}{1+\alpha^2 a^2}\,x^4 \Big) , \label{delta_x_init}
\end{align}
see Eq.~(25) of \cite{OvcharenkoPodolskyAstorino:2024}.
Here $m$~is the \emph{mass} parameter, $a$~denotes the Kerr-like \emph{rotation}, $l$~is the \emph{NUT parameter}, $\alpha$~is \emph{acceleration}, $e$~and $g$~represent \emph{electric and magnetic charges}, and $\Lambda$~is the \emph{cosmological constant}.
It is assumed that $m, a, \alpha$ are positive (or zero), while $l, e, g, \Lambda$ can take any value. These quantities have \emph{usual physical dimensions}, namely $m, a, l, e, g$ have the dimension of length, $\alpha$ and $\sqrt{\Lambda}$ are inverse length. The coordinates $t, r$ have the dimension of length, while $x, \varphi$ are dimensionless.
\newpage

In order to get the correct limits and asymptotic behavior of the general metric \eqref{ds2_simpl}, the dimensionless normalization constant ${C_f>0}$ should be chosen as
\begin{equation}
  C_f = \frac{1}{1+\alpha^2a^2} \,.\label{Cf-choice}
\end{equation}

The related \emph{electromagnetic field} has the vector potential ${A(r,x)=A_t\,\dd t + A_\varphi\,\dd\varphi}$ with
\begin{align}
 A_t =& \,\sqrt{\frac{1+\alpha^2(a^2-l^2)}{1+\alpha^2a^2}}\,\frac{-1}{\rho^2\, l}
       \Big[ \big[\,g (r + \alpha a l +  \alpha^2 a^2 r) + e l \,\big]\,r  \nonumber\\
 & \hspace{3mm} + l\,(g-\alpha a e) (a+2 \alpha l\, r+\alpha^2a\,r^2)\,x
  \label{At}\\
 & \hspace{3mm} +(a + \alpha l\,r)\big[\,ag + \alpha l(gr-el) + \alpha^2a\,[g(a^2-l^2)-el\,r] \big]\,x^2\Big]
   + \frac{g}{l} \,,  \nonumber \\
 A_\varphi =& \,\sqrt{\frac{1+\alpha^2(a^2-l^2)}{1+\alpha^2a^2}}\,\frac{x}{\rho^2}
         \Big[
    \alpha l^3 (e+\alpha ag)x  + l^2\big[\alpha^2a(e+\alpha a g)\,xr + (\alpha a e-g)\big] \label{Aphi}\\
 & \hspace{3mm}+ (1+\alpha^2a^2)\big[ ae(x+\alpha\,r)\,r - gal\,x + g(r-\alpha a^2x)\,r + el(2+\alpha xr)r \big]  \Big] - a\, A_t\,.  \nonumber
\end{align}

\subsection {Transformation in a fully general case}
\label{sc:generalPD}

Now, performing the transformation\footnote{The upper sign is used for ${a^2>l^2}$, while the lower sign is used in the case ${a^2<l^2}$ (including ${a=0}$). Therefore, $   a^2(I^2 \mp J^2) = a^2 [1+\alpha^2(a^2-l^2)]^2 + 4\alpha^2 l^2(l^2-a^2)$,
and thus
 $   2aK = a\,[1+\alpha^2(a^2-l^2)] + \sqrt{ a^2[1+\alpha^2(a^2-l^2)]^2 + 4\alpha^2 l^2(l^2-a^2)}$.}
\begin{align}
t &=\ \dfrac{a}{\alpha(a^2-l^2)}\,\sqrt{\dfrac{K-1}{\sqrt{I^2 \mp J^2}}}\,
    \Big[\big[K-\alpha^2(a^2-l^2)\big]\,\tau' + \frac{K-1}{\alpha^2(a^2-l^2)}\,\phi'\,\Big],\label{direct_transformation_A-PD-t}\\
\varphi &=\ \dfrac{1}{\alpha(a^2-l^2)}\,\sqrt{\dfrac{K-1}{\sqrt{I^2 \mp J^2}}}\,
    \Big[\phi'-\alpha^2(a^2-l^2)\,\tau'\,\Big],\label{direct_transformation_A-PD-phi}\\[2mm]
x         &=\ \frac{aK\,x'-l}{aK-l\alpha^2(a^2-l^2)\,x'}\,, \label{direct_transformation_A-PD-x}\\[3mm]
\alpha\,r &=\ \frac{aK(a^2-l^2)\,\alpha\,r'-al(K-1)}{(a^2K-l^2) - l(a^2-l^2)\,\alpha\,r'}\,, \label{direct_transformation_A-PD-r}
\end{align}
where the dimensionless constants $I$, $J$, and $K$ are defined as
\begin{align}
    I := &\ 1+\alpha^2(a^2-l^2)\, , \nonumber\\
    J := &\ 2 \alpha\,\dfrac{l}{a}\,\sqrt{|a^2-l^2|}\,, \label{defI-and -defJ}\\
    2K := &\ I + \sqrt{I^2 \mp J^2}\,, \nonumber
\end{align}
the A$^+$ metric \eqref{ds2_simpl}--\eqref{delta_x_init} is transformed to the Pleba\'nski-Demia\'nski metric (denoted as PD$_{\alpha}$ in \cite{OvcharenkoPodolskyAstorino:2024}, see also Eq.~$(16.6)$ with ${\omega=1}$ in~\cite{GriffithsPodolsky:2009}), namely
\begin{align}
    \dd s^2=\dfrac{1}{{(1-\alpha'\,r'\,x')}^{\,2}}\bigg[
    &- \dfrac{Q'}{{r'}^{\,2}+{x'}^{\,2}}\,(\dd\tau'  -  {x'}^{\,2}\, \dd\phi' )^2
     + \dfrac{P'}{{r'}^{\,2}+{x'}^{\,2}}\,(\dd\tau'  +  {r'}^{\,2}\, \dd\phi' )^2 \nonumber\\
    &+ c^2 ({r'}^{\,2}+{x'}^{\,2}) \Big(\,\dfrac{\dd {r'}^{\,2}}{Q'} + \dfrac{\dd {x'}^{\,2}}{P'}\,\Big)\bigg], \label{PD-GP-form}
\end{align}
with the dimensionless metric functions\footnote{Here the coordinates $r', x'$ are dimensionless, while $\tau', \phi'$ have the physical dimension of length.}
 \begin{equation}
 \begin{array}{l}
P'(x') = k'+2n' x' - \epsilon'x'^{\,2} +2\alpha'm' x'^{\,3}
- \big[\alpha'^{\,2} ({k'}+{e'}^2 + {g'}^2) + c^2 \Lambda/3 \big]\,x'^{\,4}   \,, \\[8pt]
Q'(r') = (k'+e'^2 + g'^2)-2m' r' + \epsilon'r'^{\,2} - 2\alpha' n' \,r'^{\,3} - (\alpha'^{\,2} k' + c^2  \Lambda/3 )\,r'^{\,4} \,.
 \end{array}
 \label{P'Q'eqns}
 \end{equation}
The constant $c^2$ is
\begin{equation}
 c^2 = \frac{\sqrt{I^2 \mp J^2}}{K-1}\, \alpha^2(a^2-l^2)^2\,C_f  \,,\label{C-definition}
\end{equation}
see Eqs.~(A57) and (A54) in our previous paper \cite{OvcharenkoPodolskyAstorino:2024}.

Actually, \eqref{direct_transformation_A-PD-t}--\eqref{defI-and -defJ} is the \emph{same} transformation which we found in \cite{OvcharenkoPodolskyAstorino:2024} in the subcase ${\Lambda=0}$, see Eqs.~(34)--(38) therein. It also changes the vector potential \eqref{At}--\eqref{Aphi} to
\begin{align}
 A(r', x')  = - \frac{e' + {\rm i}\,g' }{r' + {\rm i}\,x'}\,\big(\,\dd\tau' - {\rm i}\,r' x'\dd\phi' \,\big) .\label{A_PD-scaled}
\end{align}
Therefore, the new Astorino class \cite{Astorino:2024b} and the classic Pleba\'nski-Demia\'nski class~\cite{PlebanskiDemianski:1976} are diffeomorphic: They both represent \emph{all black-hole solutions} of the \hbox{Einstein-Maxwell-$\Lambda$} field equations of algebraic type~D with double aligned, non-null electromagnetic field.
\newpage

Moreover, the transformation \eqref{direct_transformation_A-PD-t}--\eqref{defI-and -defJ} \emph{uniquely relates
the eight constants} of the Pleba\'nski-Demia\'nski metric to the convenient Astorino seven physical parameters as
\begin{align}\label{direct_transformation_A-PD_parameters-simplified}
\alpha' &=\ \alpha\,a\,\big[\,K - \alpha^2(a^2-l^2)\big]\,, \nonumber\\
k'  &=\ \frac{1}{I^2 \mp J^2}\,\,\frac{K-1}{\alpha^2(a^2-l^2)}\,\,I\, L
      + k'_\Lambda\,, \nonumber\\
n'  &=\ \frac{-1}{I^2 \mp J^2}\,
      \Big[\,\frac{K-1}{\alpha^2(a^2-l^2)}\,\,I\, M  - \big[1-\alpha^2(a^2-l^2)\big]\,L\,\frac{l}{a}\, \Big]
      + n'_\Lambda, \nonumber\\
\epsilon'  &=\ \frac{1}{I^2 \mp J^2}\, \big[1-\alpha^2(a^2-l^2)\big]
      \Big( I\, L  + 4\,M\,\frac{l}{a} \Big)
      - (e^2+g^2)\frac{K-1}{(a^2-l^2)}\,\frac{I}{\sqrt{I^2 \mp J^2}}
      + \epsilon'_\Lambda\,, \nonumber\\
\alpha' m' &=\ \frac{1}{I^2 \mp J^2}\, \Big[ \,
    \big(K-1\big)\,I\, M
       \\
&      \hspace{6.5mm} + \big[1-\alpha^2(a^2-l^2)\big]\Big( I\, M
       + \alpha^2 \big[(a^2-l^2)\,L+ (e^2+g^2)\sqrt{I^2 \mp J^2}\,\big]\frac{l}{a} \Big)\Big]
       + \alpha' m'_\Lambda, \nonumber\\
{e'}^2 + {g'}^2 &=\ ({e}^2 + {g}^2)\, \frac{K-1}{\alpha^2(a^2-l^2)^2}\,\frac{I}{\sqrt{I^2 \mp J^2}} \,,
        \nonumber
\end{align}
where the specific dimensionless combinations of the physical parameters are
\begin{align}
    L := &\ I+2\alpha\,m\,\frac{l}{a} + \alpha^2(e^2+g^2)\frac{1}{K}\,\frac{l^2}{a^2}\, , \label{defL}\\
    M := &\ \alpha\,m\,I+\alpha^2(2a^2-2l^2+e^2+g^2)\frac{l}{a}\,. \label{defM}
\end{align}
In \eqref{defL} we can alternatively use the identity
\begin{align}
\frac{1}{K}\,\frac{l^2}{a^2} = \frac{I-\sqrt{{I^2 \mp J^2}}}{2\alpha^2(a^2-l^2)}\,.
\end{align}
The additional terms in \eqref{direct_transformation_A-PD_parameters-simplified} depending on the cosmological constant $\Lambda$ are
\begin{align} \label{coeff_Lambda}
 k'_{\Lambda}&=-\frac{\Lambda\,l^4}{I^2 \mp J^2}
    \dfrac{a^2(3K-1)(K-1)+\alpha^2(a^2-l^2)^2}{3a^4K^2(1+\alpha^2 a^2)}\,, \nonumber\\[2mm]
 n'_{\Lambda}&=-\frac{\Lambda\,l^3}{I^2 \mp J^2}
    \dfrac{2\alpha^2(a^2-l^2)l^2a - aK[2a^2+\alpha^2(a^2-l^2)(2a^2+l^2)]
    +3a^3K^2(2-K)}{3a^4K^2(1+\alpha^2 a^2)}\,, \nonumber\\[2mm]
 \epsilon'_{\Lambda}&=\frac{\Lambda\,l^2}{I^2 \mp J^2} \times  \\
 &\hspace{6mm}
    \dfrac{\alpha^4(a^2-l^2)^2l^4-4\alpha^2(a^2-l^2)l^2a^2K+2a^2K^2[a^2+\alpha^2(a^4-l^4)]+a^4K^3(K-4)}
        {a^4K^2(1+\alpha^2 a^2)}\,, \nonumber\\
 \alpha' m'_{\Lambda}&=\dfrac{\Lambda\,l}{I^2 \mp J^2} \times \nonumber\\
 &\hspace{6mm}
    \dfrac{3\alpha^4(a^2-l^2)^2l^4-6\alpha^2(a^2-l^2)l^2a^2K+a^2K^2[2a^2+\alpha^2(a^2-l^2)(2a^2+l^2)]-2a^4K^3}
     {3a^3 K (1+\alpha^2a^2)}\,. \nonumber
\end{align}
All of them are \emph{linear} in $\Lambda$. Actually, they are proportional to $\Lambda\,l$, i.e., vanish when ${\Lambda=0}$ and/or when ${l=0}$.

\subsection {The special case ${\Lambda=0}$: no cosmological constant}
\label{sc:lambda=0PD}

For ${\Lambda=0}$, the expressions \eqref{direct_transformation_A-PD_parameters-simplified} reduce to the forms presented in Eqs.~(44) of \cite{OvcharenkoPodolskyAstorino:2024}. In such a case, both metric functions $P'$ and $Q'$ given by \eqref{P'Q'eqns} can be written in the factorized form as the product of two quadratic expressions, namely
\begin{align}
 P'=\dfrac{1}{I^2 \mp J^2}&
    \Bigg[ \dfrac{I(K-1)}{\alpha^2(a^2-l^2)}+2\,[1-\alpha^2(a^2-l^2)]\dfrac{l}{a}\,x'-I\,[K-\alpha^2(a^2-l^2)]\,x'^{\,2}\Bigg]\nonumber\\
    \times& \Bigg[ L -2M\,x' + \alpha^2\,\big[(a^2-l^2)L+(e^2+g^2)\sqrt{I^2 \mp J^2}\,\big]\,x'^{\,2}\Bigg], \label{P'-factorized}\\
 Q'=\dfrac{1}{I^2 \mp J^2}&
    \Bigg[I-2\alpha l\,[1-\alpha^2(a^2-l^2)]\,r'-\alpha^2(a^2-l^2)I\,r'^{\,2}\Bigg]\nonumber\\
    \times& \Bigg[ \dfrac{K-1}{\alpha^2(a^2-l^2)^2}\,\big[(a^2-l^2)L + (e^2+g^2)\sqrt{I^2 \mp J^2}\,\big] \nonumber
     \\& -\dfrac{2}{\alpha a}M\,r'+L\,[K-\alpha^2(a^2-l^2)]\,r'^{\,2}\Bigg], \label{Q'-factorized}
\end{align}
see Eqs.~(48) and~(49) in our previous work \cite{OvcharenkoPodolskyAstorino:2024}. It also involves a detailed discussion of the case ${\Lambda=0}$.

In the general case ${\Lambda\ne0}$, such a simultaneous factorization is not possible. However, there is a considerable simplification for subcases of the Pleba\'nski-Demia\'nski PD$_{\alpha}$ metric \eqref{PD-GP-form}
when (some of) the physical parameters vanish, as we will now describe.

\subsection {The special case ${l=0}$: no NUT}
\label{sc:l=0PD}

For black holes without the NUT twist (that is for ${l=0}$), the dimensionless constants \eqref{defI-and -defJ},
\eqref{defL}, \eqref{defM} reduce to
\begin{align} \label{defI-and -defJ-l=0}
    I = 1+\alpha^2a^2 \,, \qquad
    J = 0 \,,\qquad
    K = I \,,\qquad
    L = I\,,\qquad
    M = \alpha\,m\,I\,,
\end{align}
and also all additional coefficients \eqref{coeff_Lambda} proportional to $\Lambda$ vanish.

The Pleba\'nski-Demia\'nski parameters \eqref{direct_transformation_A-PD_parameters-simplified} thus simplify to
\begin{align}\label{direct_transformation_A-PD_parameters-l=0}
\alpha' &= \alpha\,a \,, \nonumber\\
k'  &= 1 \,, \nonumber\\
n'  &= -\alpha\,m \,,\\
\epsilon'  &= 1 - \alpha^2(a^2+e^2+g^2)\,, \nonumber\\
\alpha' m' &= \alpha\,m \,,\nonumber\\
({e'}^2 + {g'}^2) &= ({e}^2 + {g}^2)/a^{2}\,, \nonumber
\end{align}
and the constant \eqref{C-definition} with \eqref{Cf-choice} is
\begin{equation}
  c^2 = a^2 \,.\label{C-for-l=0}
\end{equation}
The two key metric functions thus reduce to
\begin{align}
 P' &=       \big( 1 - x'^{\,2} \big)\big( 1 -2\alpha\,m\,x' + \alpha^2 (a^2+e^2+g^2)\,x'^{\,2}\big)
      - \frac{\Lambda}{3}\,a^2 x'^{\,4}, \label{P'-factorized:l=0}\\
 Q' &=a^{-2} \big( 1 -\alpha^2a^2 r'^{\,2} \big) \big( a^2+e^2+g^2 - 2\,m\,a\,r' + a^2 r'^{\,2} \big)
      - \frac{\Lambda}{3}\,a^2 r'^{\,4}. \label{Q'-factorized:l=0}
\end{align}

Recall that the Astorino parameters in \eqref{P'-factorized:l=0}, \eqref{Q'-factorized:l=0} have the proper physical dimensions, while the PD coordinates $r', x'$ are dimensionless. Introducing a coordinate $\bar{r}$ of dimension length and a dimensionless coordinate $\bar{\phi}$ by the convenient rescalings
\begin{align} \label{barr-baarphi}
    \bar{r} := a\, r'\,,\qquad  \bar{\phi} := \phi'/a\,,
\end{align}
the Pleba\'nski-Demia\'nski PD$_{\alpha}$ metric \eqref{PD-GP-form} becomes
\begin{align}
    \dd s^2=\dfrac{1}{{(1-\alpha\,\bar{r}\,x')}^{\,2}}\bigg[
    &- \dfrac{\bar{Q}}{\bar{r}^{\,2}+a^2{x'}^{\,2}}\,(\dd\tau'  -  a\,{x'}^{\,2}\, \dd \bar{\phi}\,)^2
     + \dfrac{P'}{\bar{r}^{\,2}+a^2{x'}^{\,2}}\,(a\,\dd\tau'  +  \bar{r}^{\,2}\, \dd \bar{\phi}\,)^2 \nonumber\\
    &+ (\bar{r}^{\,2}+a^2{x'}^{\,2}) \Big(\,\dfrac{\dd \bar{r}^{\,2}}{\bar{Q}} + \dfrac{\dd {x'}^{\,2}}{P'}\,\Big)\bigg], \label{PD-GP-form-Astor}
\end{align}
where ${\bar{Q} := a^2\, Q'}$ has the form
\begin{align}
 \bar{Q}(\bar{r}) &= \big( 1 -\alpha^2\,\bar{r}^{\,2} \big) \big( a^2+e^2+g^2 - 2\,m\,\bar{r} + \bar{r}^{\,2} \big)      - \frac{\Lambda}{3}\,\bar{r}^{\,4}. \label{Q'-factorized:l=0-rbar}
\end{align}
In this metric it is possible to set any of the remaining six physical parameters
${\alpha, a, e, g, m, \Lambda}$ to zero, in an arbitrary order. In particular, for vanishing acceleration (${\alpha=0}$) and Kerr-like rotation (${a=0}$), by introducing ${x'=\cos\theta}$ we recover
\begin{align}
    \dd s^2 =
    &- f(\bar{r})\,\dd\tau'^{\,2}+ \dfrac{\dd \bar{r}^{\,2}}{f(\bar{r})}
     + \bar{r}^{\,2} \big( \dd \theta^{2} +  \sin^2 \theta \, \dd \bar{\phi}^{\,2} \big), \label{PD-GP-form-alpha=0=a}
\end{align}
where
\begin{align}
 f(\bar{r}) &=
   1 - \dfrac{2\,m}{\bar{r}} + \dfrac{e^2+g^2}{\bar{r}^{\,2}} - \frac{\Lambda}{3}\,\bar{r}^{\,2}, \label{f:l=0}
\end{align}
which is the usual metric of the Reissner-Nordstr\"om-(anti-)de~Sitter black hole, see~\cite{GriffithsPodolsky:2009}.

\subsection {The special case ${\alpha=0}$: no acceleration}
\label{sc:alpha=0PD}

Non-accelerating black holes are obtained in the limit ${\alpha\to0}$. In such a case,
\begin{align} \label{defI-and -defJ-alpha=0}
    I = 1 \,, \qquad
    J = 0 \,,\qquad
    K = 1 \,,\qquad
    L = 1 \,,\qquad
    M = \alpha \,m \to 0\,.
\end{align}
In fact, ${K-1 = \alpha^2 \,(a^2-l^2)^2/a^2 }$
which enables us to evaluate the parameters \eqref{direct_transformation_A-PD_parameters-simplified}, \eqref{coeff_Lambda} as
\begin{align}\label{direct_transformation_A-PD_parameters-alpha=0}
\alpha' &= 0 \,, \nonumber\\
k'  &= 1 - \frac{l^2}{a^2} \,, \nonumber\\
n'  &= \frac{l}{a}\Big( 1 -\frac{\Lambda}{3}\,l^2 \Big)\,,\\
\epsilon'  &= 1 - \Lambda\,l^2 \,  , \nonumber\\
m' &= m/a \, ,\nonumber\\
{e'}^2 + {g'}^2 &= ({e}^2 + {g}^2)/a^2 \,. \nonumber
\end{align}

The metric functions  \eqref{P'Q'eqns} thus reduce to
\begin{align}
 P' &= 1 - \Big(\,x' - \dfrac{l}{a}\,\Big)^2 \Big[\,1 + \frac{\Lambda}{3}\,a^2 x' \Big(x'+2\frac{l}{a}\Big)\Big], \label{P'-factorized:alpha=0}\\[1mm]
 Q' &= a^{-2} \Big( a^2-l^2+e^2+g^2 - 2\,m\,a\,r' + (1-\Lambda\,l^2)\, a^2 r'^{\,2}
     -\frac{\Lambda}{3}\, a^4 r'^{\,4}\Big), \label{Q'-factorized:alpha=0}
\end{align}
and \eqref{C-definition} with \eqref{Cf-choice} in this case give
\begin{equation}
 c^2 = a^2 \,.\label{C-for-alpha=0}
\end{equation}
Performing the transformation \eqref{barr-baarphi}, that is ${\bar{r} := a\, r'}$, ${\bar{\phi} := \phi'/a}$,
 with a rescaling ${\bar{Q} := a^2\, Q'}$, and an additional shift
\begin{align} \label{barx}
    \bar{x} := x' - l/a\,,\
\end{align}
the Pleba\'nski-Demia\'nski metric \eqref{PD-GP-form} for ${\alpha=0}$ takes the form
\begin{align}
    \dd s^2 =
    &- \dfrac{\bar{Q}}{\bar{r}^{\,2}+(l+a\,\bar{x})^2}\,\big(\dd\tau'  -  (l+a\,\bar{x})^2 \,\dd \bar{\phi}/a \,\big)^2
     + \dfrac{\bar{P}}{\bar{r}^{\,2}+(l+a\,\bar{x})^2}\,\big(a\,\dd\tau'  +  \bar{r}^{\,2}\, \dd \bar{\phi}\,\big)^2 \nonumber\\
    &+ \big[\bar{r}^{\,2}+(l+a\,\bar{x})^2\big] \Big(\,\dfrac{\dd \bar{r}^{\,2}}{\bar{Q}} + \dfrac{\dd \bar{x}^{\,2}}{\bar{P}}\,\Big), \label{PD-GP-form-Astor-alpha=0}
\end{align}
where
\begin{align}
 \bar{P}(\bar{x}) &= (1 - \bar{x}^{\,2})
     - \frac{\Lambda}{3}\,(l+a\,\bar{x})(3l+a\,\bar{x})\,\bar{x}^{\,2}, \label{P'-factorized:alpha=0-xbar}\\
 \bar{Q}(\bar{r}) &= a^2-l^2+e^2+g^2 - 2\,m\,\bar{r} + (1-\Lambda\,l^2)\, \bar{r}^{\,2}
     -\frac{\Lambda}{3}\, \bar{r}^{\,4}. \label{Q'-factorized:alpha=0-rbar}
\end{align}

In particular, for ${l=0}$ it simplifies to
\begin{align}
    \dd s^2 =
    &- \dfrac{\bar{Q}}{\bar{r}^{\,2}+a^2\bar{x}^2}\,\big(\dd\tau'  -  a\,\bar{x}^2 \,\dd \bar{\phi}\,\big)^2
     + \dfrac{\bar{P}}{\bar{r}^{\,2}+a^2\bar{x}^2}\,\big(a\,\dd\tau'  +  \bar{r}^{\,2}\, \dd \bar{\phi}\,\big)^2 \nonumber\\
    &+ \big(\bar{r}^{\,2}+a^2\bar{x}^2\big) \Big(\,\dfrac{\dd \bar{r}^{\,2}}{\bar{Q}} + \dfrac{\dd \bar{x}^{\,2}}{\bar{P}}\,\Big), \label{PD-GP-form-Astor-alpha=0-and-l=0}
\end{align}
with
\begin{align}
 \bar{P}(\bar{x}) &= (1 - \bar{x}^{\,2}) - \frac{\Lambda}{3} a^2 \bar{x}^{\,4}\,, \label{P'-factorized:alpha=0-xbar-and-l=0}\\
 \bar{Q}(\bar{r}) &= a^2+e^2+g^2 - 2\,m\,\bar{r} +  \bar{r}^{\,2}
     -\frac{\Lambda}{3}\, \bar{r}^{\,4}\,, \label{Q'-factorized:l=0-rbar-and-l=0}
\end{align}
which is the PD form of the Kerr-Newman-(anti-)de~Sitter black hole.

\subsection {The special case ${a=0}$: no Kerr-like rotation}
\label{sc:a=0PD}

For black holes without the Kerr-like rotation, by setting ${a = 0}$
we get
\begin{align} \label{defI-and -defJ-a=0}
    I &=  1-\alpha^2l^2 \,, \qquad
    a^2(I^2 \mp J^2) = 4\alpha^2l^4 \,,\qquad
    aK = \alpha\,l^2   \,,  \nonumber\\
    aL&= \alpha\,\big(2ml + e^2+g^2\big)\,,\qquad
    aM = \alpha^2l\,\big(-2l^2 +e^2+g^2\big)\,,
\end{align}
so that
\begin{align}\label{direct_transformation_A-PD_parameters-a=0}
\alpha' &= \alpha^2 \,l^2 \,, \nonumber\\
k'  &= \frac{\alpha^2 l^2-1}{4\alpha^2 l^4}\,\big(2m\,l +e^2+g^2   \big) - \dfrac{\Lambda}{3\alpha^2} \,, \nonumber\\
n'  &= \frac{(m-l)\,l+(e^2+g^2)}{2 \alpha\, l^3} + \frac{\alpha}{2}(m+l) + \dfrac{\Lambda l}{6 \alpha}\,,\\
\epsilon'  &= -2 (1+\alpha^2l^2) + \frac{1}{2}(e^2+g^2)\Big( \alpha^2 + \frac{3}{l^2}  \Big)  , \nonumber\\
\alpha' m' &= \frac{\alpha}{2}\,\Big[ -(m+l) + \alpha^2l^2(l-m)
    +\frac{e^2+g^2}{l} + \dfrac{\Lambda}{3}\,l^3 \,\Big] ,\nonumber\\
{e'}^2 + {g'}^2 &= \frac{1-\alpha^2l^2}{2\alpha^2 l^4}\,({e}^2 + {g}^2)\,, \nonumber
\end{align}
Also, \eqref{C-definition} with \eqref{Cf-choice} yields
\begin{equation}
 c^2 = 2\alpha^2\l^4 \,.\label{C-for-a=0}
\end{equation}

Therefore, the two key PD metric functions \eqref{P'Q'eqns} are explicitly
\begin{align}
 P'=\dfrac{1}{4\alpha^2l^4}&
    \Big[ ( 1 - \alpha\,l\, x')^2 - \alpha^2l^2( 1 + \alpha\,l\, x')^2 \Big]\nonumber\\
    \times& \Big[ - \big(2ml + e^2+g^2\big)
    + 2 \alpha l\,(-2l^2 +e^2+g^2)\,x'
    - \alpha^2l^2(-2ml +e^2+g^2)\,x'^{\,2}\Big] \nonumber\\
    & \hspace{-10.5mm}
    -\dfrac{\Lambda}{3\alpha^2}\,( 1 - \alpha\,l\, x')^2\,( 1 + \alpha\,l\,x' + \alpha^2l^2 x'^{\,2} ),
 \label{P'-factorized:a=0}\\[8pt]
 Q'=\dfrac{1}{4\alpha^2l^4}&
    \Big[ ( 1 - \alpha\,l\, r')^2 - \alpha^2l^2( 1 + \alpha\,l\, r')^2 \Big]\nonumber\\
    \times& \Big[ (-2ml+e^2+g^2) -2 \alpha l\,(-2l^2 +e^2+g^2)\,r' + \alpha^2l^2(2ml + e^2+g^2)\,r'^{\,2}\Big] \nonumber\\
    & \hspace{-10.5mm}
   -\dfrac{\Lambda}{3\alpha^2}\,( 1 + \alpha\,l\, r' )^2\,( 1 - \alpha\,l\,r' + \alpha^2l^2 r'^{\,2} ).
    \label{Q'-factorized:a=0}
\end{align}
in the line element \eqref{PD-GP-form},
\begin{align}
    \dd s^2=\dfrac{1}{{(1-\alpha^2 l^2\,r'\,x')}^{\,2}}\bigg[
    &- \dfrac{Q'}{{r'}^{\,2}+{x'}^{\,2}}\,(\dd\tau'  -  {x'}^{\,2}\, \dd\phi' )^2
     + \dfrac{P'}{{r'}^{\,2}+{x'}^{\,2}}\,(\dd\tau'  +  {r'}^{\,2}\, \dd\phi' )^2 \nonumber\\
    &+ 2\alpha^2\l^4  ({r'}^{\,2}+{x'}^{\,2}) \Big(\,\dfrac{\dd {r'}^{\,2}}{Q'} + \dfrac{\dd {x'}^{\,2}}{P'}\,\Big)\bigg]. \label{PD-GP-form-a=0}
\end{align}
This form of the metric, expressed in terms of the Astorino physical parameters, identifies the large elusive family of accelerating NUTty black holes \emph{without the Kerr-like rotation} within the Pleba\'nski-Demia\'nski family of metrics, which now also include electric and magnetic charges, and any value of the cosmological constant.\\

Notice also that, analogously to  \eqref{barr-baarphi}, we can re-introduce the proper physical dimension to the dimensionless coordinate $r'$ by
\begin{align} \label{barr-baarphi-l}
    \bar{r} := \alpha\,l^2\, r'  \,,
\end{align}
so that the PD metric function $Q'$ becomes ${\bar{Q} := \alpha^2 l^4\, Q'}$, namely
\begin{align}
 \bar{Q}(\bar{r}) = \dfrac{1}{4}&
    \Big[ \,\frac{1}{l^2} ( l - \bar{r} )^2 - \alpha^2 ( l + \bar{r})^2 \Big] \nonumber\\
    \times& \Big[ (-2ml+e^2+g^2) -2 \,(-2l^2 +e^2+g^2)\frac{\bar{r}}{l}
    + (2ml + e^2+g^2)\frac{\bar{r}^{\,2}}{l^2}\,\Big] \nonumber\\
    & \hspace{-4.5mm} -\dfrac{\Lambda}{3}\, ( l + \bar{r})^2\,( l^2 - l\,\bar{r} + \bar{r}^{\,2} )\,.
    \label{Q'-factorized:l=0-rbar-l}
\end{align}
We will analyze this spacetime in more detail in Section~\ref{sc:a=0GP}, including its special subcases corresponding to the limits ${ \alpha \to 0}$ and ${l \to 0}$.

\newpage

\section{Transformation to the Griffiths-Podolsk\'y metric}
\label{sc:GP-form}

Now we will relate the PD$_{\alpha}$ form \eqref{PD-GP-form} of the type~D black holes to the Griffiths-Podolsk\'y (GP) form of these solutions, summarized in \cite{GriffithsPodolsky:2009}. In particular, it will clarify the relation between the convenient Astorino  parameters $\alpha,~a,~l,~m,~e,~g$ and the physical parameters $\tilde{\alpha},~\tilde{a},~\tilde{l},~\tilde{m},~\tilde{e},~\tilde{g}$ employed in the previous GP form of the metric \cite{GriffithsPodolsky:2005, GriffithsPodolsky:2006, PodolskyGriffiths:2006}.

\subsection {Transformation in a fully general case}
\label{sc:generalGP}

Following these works, we perform a coordinate transformation (applied also in \cite{OvcharenkoPodolskyAstorino:2024})
\begin{equation}
    x'= \tilde{l} + \tilde{a}\,\tilde{x},\qquad
 \tau'=t-\dfrac{(\tilde{a}+\tilde{l})^2}{\tilde{a}}\,\varphi,\qquad
 \phi'=-\dfrac{1}{\tilde{a}}\,\varphi,\qquad
    r'=\tilde{r},\label{coord_tr_2}
\end{equation}
where $\tilde{a}$ represents the Kerr-like rotational parameter, while $\tilde{l}$ represent the NUT-like parameter. After these linear transformations, the PD$_{\alpha}$ metric \eqref{PD-GP-form} becomes
\begin{align}
    \dd s^2=\dfrac{1}{{\tilde{\Omega}}^2}
    \Bigg[&-\dfrac{\tilde{\mathcal{Q}}}{\,\tilde{\rho}^{\,2}}
    \Big[\dd t-\big(\tilde{a}(1-\tilde{x}^2)+2\tilde{l}(1-\tilde{x})\big)\dd\varphi\Big]^2
     + c^2 \dfrac{\,\tilde{\rho}^{\,2}}{\tilde{\mathcal{Q}}}\,\dd\tilde{r}^2\nonumber\\
    &+ c^2 \dfrac{\,\tilde{\rho}^{\,2}}{\tilde{\mathcal{P}}}\,\dd\tilde{x}^2
     +\dfrac{\tilde{\mathcal{P}}}{\,\tilde{\rho}^{\,2}}
    \big[\tilde{a}\,\dd t-\big(\tilde{r}^2+(\tilde{a}+\tilde{l})^2\big)\,\dd\varphi\big]^2\Bigg],
    \label{ds2_accel_kerr_new}
\end{align}
in which the constant $c^2$ is defined in \eqref{C-definition},
\begin{align}
\tilde{\Omega} &:= 1 - \tilde{\alpha}\,\tilde{r}\,(\tilde{l}+\tilde{a}\,\tilde{x}), \\[2mm]
\tilde{\rho}^{\,2} &:= \tilde{r}^{\,2}+(\tilde{l}+\tilde{a}\,\tilde{x})^2, \\[2mm]
\tilde{\mathcal{P}}(\tilde{x}) &:= \dfrac{1}{\tilde{a}^2}\,P'(\tilde{l}+\tilde{a}\,\tilde{x}),
\label{mathcalP}\\[1mm]
\tilde{\mathcal{Q}}(\tilde{r}) &:= Q' (\tilde{r})\,,
\label{mathcalQ}
\end{align}
where
\begin{equation}
\tilde{\alpha} = \alpha' \,,
\label{acceleration}
\end{equation}
and the functions $P'(x'), Q'(r')$ are given by expressions \eqref{P'Q'eqns} with \eqref{direct_transformation_A-PD_parameters-simplified}, \eqref{coeff_Lambda}. It is the general Griffiths-Podolsk\'y form of the metric (with ${\omega=1}$), as summarized in Eq.~(16.12) in~\cite{GriffithsPodolsky:2009}.

In particular, because the \emph{GP acceleration parameter $\tilde{\alpha}$ is the same as the PD acceleration parameter $\alpha'$}, see \eqref{acceleration}, using \eqref{direct_transformation_A-PD_parameters-simplified} we can relate the GP acceleration to the  A parameters as
\begin{equation}
    \tilde{\alpha} = \alpha\,a\,\big[\,K - \alpha^2(a^2-l^2)\big],
    \label{tild_alph}
\end{equation}
where $K$ is given by \eqref{defI-and -defJ}. Expressed explicitly, this is actually a quite complicated relation
\begin{equation}\label{direct_transformation_A-PD-acceleration-explicit-again}
 \tilde{\alpha} = \alpha\,\,\tfrac{1}{2}\Big[\,a - \alpha^2 a (a^2-l^2)
+ \sqrt{a^2+\alpha^4 a^2(a^2-l^2)^2+2\alpha^2(a^2-l^2)(a^2-2l^2)}\, \Big].
\end{equation}
Thus, ${\alpha=0}$ implies ${\tilde{\alpha}=0}$. By setting ${l=0}$ we get ${\tilde{\alpha}=\alpha\,a}$, and for ${a=0}$ we get~${ \tilde{\alpha}=\alpha^2 l^2}$. It means that the GP acceleration parameter $\tilde{\alpha}$ \emph{also vanishes for} ${a=0=l}$. This particular degeneracy is an unfortunate feature of the original GP representation of the class of type~D black holes.

Having determined the acceleration parameter $\tilde{\alpha}$, it is now necessary to express the GP ``rotational'' parameters $\tilde{a}$ and $\tilde{l}$ in terms of the Astorino parameters of the A$^+$~metric \eqref{ds2_simpl}. To this end, we will employ the relation between the PD and  GP parameters derived already in the original works \cite{GriffithsPodolsky:2005, GriffithsPodolsky:2006, PodolskyGriffiths:2006}, summarized in \cite{GriffithsPodolsky:2009}.
It uniquely follows from the requirement that the function $\tilde{\mathcal{P}}$ in \eqref{ds2_accel_kerr_new}, which is a \emph{quartic polynomial in} $\tilde{x}$,
can be written in the specific factorized form\footnote{Applying the transformation ${\tilde{x}=\cos\theta}$ to spherical-like coordinate $\theta$, the GP metric is finally obtained.}
\begin{equation}
\tilde{\mathcal{P}}(\tilde{x})=(1-\tilde{x}^2)(a_0-a_3\,\tilde{x}-a_4\,\tilde{x}^2).\label{tild_P_simpl}
\end{equation}
Generally, ${\tilde{\mathcal{P}}(\tilde{x}) = a_0+a_1\,\tilde{x}+a_2\,\tilde{x}^2+a_3\,\tilde{x}^3+a_4\,\tilde{x}^4}$,
where
\begin{align}
a_0 = \tilde{a}^{\,-2} A_0\,,\qquad
a_1 = 2 \tilde{a}^{\,-1} A_1\,,\qquad
a_2 = A_2\,, \qquad
a_3 = 2\tilde{a}\,A_3\,,\qquad
a_4 = \tilde{a}^2\,A_4\,,  \label{ai}
\end{align}
with
\begin{align}
A_0 &= k'+2n'\,\tilde{l}-\epsilon'\,\tilde{l}^{\,2} +2\tilde{\alpha}\,m'\, \tilde{l}^{\,3} - \lambda\,\tilde{l}^{\,4}\,,\nonumber\\
A_1 &= n' - \epsilon'\,\tilde{l} + 3\tilde{\alpha}\,m'\, \tilde{l}^{\,2} - 2\lambda\,\tilde{l}^{\,3}\,,\nonumber\\
A_2 &= -\epsilon' +  6\tilde{\alpha}\, m'\,\tilde{l} - 6 \lambda\,\tilde{l}^{\,2}\,, \label{Ai}\\
A_3 &= \tilde{\alpha}\, m' - 2\lambda\,\tilde{l}\,,\nonumber\\
A_4 &= - \lambda\,\,,  \nonumber
\end{align}
and
\begin{equation}
\lambda := \tilde{\alpha}^{2}(k'+e'^{\,2}+g'^{\,2})+c^2 \Lambda/3 \,,\label{definintion of lambda}
\end{equation}
see Eq.~(16.13) in \cite{GriffithsPodolsky:2009} for the choice ${\omega=1}$. The required factorization \eqref{tild_P_simpl} thus leads to two constraints, namely
\begin{align}
    a_1+a_3     &= 0\,, \label{constraint-1}\\
    a_0+a_2+a_4 &= 0\,. \label{constraint-2}
\end{align}

In previous works \cite{GriffithsPodolsky:2005, GriffithsPodolsky:2006, PodolskyGriffiths:2006} these were employed to explicitly express the PD parameters $\epsilon'$,~$n'$ and~$k'$ in terms of the GP ``rotational'' parameters $\tilde{a}$, $\tilde{l}$ (and the other parameters), see  Eqs.~(16.15)--(16.17) in \cite{GriffithsPodolsky:2009}, namely
\begin{align}
\epsilon'&= \dfrac{k'}{\tilde{a}^2 -\tilde{l}^{2}}+4\tilde{\alpha}\,m'\, \tilde{l} - \lambda\,(\tilde{a}^2+3\tilde{l}^{2})\,,
      \label{GP_eps_rel-book}\\[1mm]
n'       &= \dfrac{k'}{\tilde{a}^2-\tilde{l}^2}\,\tilde{l} - \tilde{\alpha}\,m' (\tilde{a}^2-\tilde{l}^2)
              + \lambda\,(\tilde{a}^2-\tilde{l}^{2})\,\tilde{l}\,, \label{GP_n_rel-book}\\[1mm]
\dfrac{k'}{\tilde{a}^2-\tilde{l}^2}&= \,  a_0 + 2\tilde{\alpha}\,m'\, \tilde{l} - 3\lambda\,\tilde{l}^{2}\,.
      \label{GP_k_rel-book}
\end{align}

However, here we need to solve the \emph{opposite problem}, which is to express  $\tilde{a}$, $\tilde{l}$ in terms of the PD parameters $\epsilon'$,~$n'$,~$k'$. By inspecting the system \eqref{GP_eps_rel-book}--\eqref{GP_k_rel-book} we observe, that it leads to equations for $\tilde{a}^2$ and $\tilde{l}$ which are \emph{only polynomial}, but they are of high order.

Actually, from the condition \eqref{constraint-1} we obtain a simple expression for $\tilde{a}^2$, namely
\begin{equation}
\tilde{a}^2 = - \frac{A_1}{A_3} =
      \frac{ \epsilon'\,\tilde{l} - n' - 3\tilde{\alpha}\,m'\, \tilde{l}^{\,2} + 2\lambda\,\tilde{l}^{\,3} }
      {\tilde{\alpha}\, m' - 2\lambda\,\tilde{l}}\,.\label{a^2-explicitly}
\end{equation}
It remains to find the expression for $\tilde{l}$. This is obtained from the constraint \eqref{constraint-2} with  \eqref{ai}, after substituting \eqref{a^2-explicitly} for $\tilde{a}^2$. It yields a complicated polynomial equation of the 6th order. Surprisingly, \emph{it can be explicitly solved}. Indeed, by introducing a dimensionless parameter
\begin{align}
    q := [(\tilde{\alpha}\,m'-2\lambda\,\tilde{l})^2-\tilde{\alpha}^2m'^{\,2}]/\lambda\,,
\end{align}
so that ${\tilde{l} = (\tilde{\alpha}\, m' \pm \sqrt{\tilde{\alpha}^2 m'^{\,2}+\lambda\,q}\,)/(2 \lambda)}$,  the equation for $q$ takes the form
\begin{align}
    q^3+2\epsilon' q^2 + \big[4(\tilde{\alpha}^2m'n'+\lambda\,k')+\epsilon'^2\big] q
        +4\big[\tilde{\alpha}^2 m'(\tilde{\alpha}^2 m' k' +n'\epsilon') - \lambda\,n'^{\,2} \big] = 0 \,.
\end{align}
Such cubic equation can be solved by Cardano's formula, leading to a general solution
\begin{align}
    \tilde{l}= \dfrac{ \tilde{\alpha}\, m' \pm \kappa}{2\,\lambda}\,, \label{l-explicitly}
\end{align}
where
\begin{equation}
    \kappa^2 = \tilde{\alpha}^2 m'^{\,2} - \frac{\lambda}{3}\,
    \Big[2\epsilon' + {\rm e}^{{\rm i}\,(2\pi/3)j}\,c_1 \Big(c_2+\sqrt{c_1^3+c_2^2}\,\,\Big)^{-1/3}
    - {\rm e}^{-{\rm i}\,(2\pi/3)j} \Big(c_2+\sqrt{c_1^3+c_2^2}\,\,\Big)^{1/3}\,\Big],\label{kappa}
\end{equation}
in which ${j=0,1,2}$, and
\begin{align}
    c_1=&\ 12\,( \lambda\,k' + \tilde{\alpha}\,m'n')-\epsilon'^{\,2}\,, \label{c1}\\
    c_2=&\ \epsilon'^{\,3}+18\, \lambda\,(2\epsilon'k' + 3n'^2)
         -18\,\tilde{\alpha}\,m'\,(\epsilon'n' + 3\tilde{\alpha}\,m'\,k')\,. \label{c2}
\end{align}

\newpage

The formulae \eqref{l-explicitly} and \eqref{a^2-explicitly} for the GP ``rotational'' parameters $\tilde{l}$ and $\tilde{a}$, respectively, are thus fully explicitly expressed in terms of the Astorino parameters. However, among the 6 possible roots of $\tilde{l}$ (some of which are complex, and thus unphysical) we have to chose those which were already identified  by us in our previous work \cite{OvcharenkoPodolskyAstorino:2024} investigating spacetimes with vanishing cosmological constant ${\Lambda=0}$. In order to get a further insight, it is useful to consider the special cases ${\Lambda=0}$, ${l=0}$, ${\alpha=0}$, ${a=0}$, demonstrating the full compatibility with the results presented in \cite{OvcharenkoPodolskyAstorino:2024}. This will be done in the following subsections \eqref{sc:lambda=0GP}--\eqref{sc:a=0GP}.\\

To finish the construction of the general Griffiths-Podolsk\'y metric form of black holes of algebraic type~D with $\Lambda$, it remains to apply a simple transformation ${\tilde{x}=\cos\theta}$ to spherical-like (angular-type) coordinate $\theta$. From \eqref{ds2_accel_kerr_new} we thus obtain the GP metric
\begin{align}
    \dd s^2=\dfrac{1}{{\tilde{\Omega}}^2}
    \Bigg[&-\dfrac{\tilde{Q}}{\,\tilde{\rho}^{\,2}}
    \Big[\dd t-\big(\tilde{a}\sin^2\theta+2\tilde{l}(1-\cos\theta)\big)\dd\varphi\Big]^2
     +c^2\dfrac{\,\tilde{\rho}^{\,2}}{\tilde{Q}}\,\dd\tilde{r}^2\nonumber\\
    &+c^2\dfrac{\,\tilde{\rho}^{\,2}}{\tilde{P}}\,\dd\theta^2
     +\dfrac{\tilde{P}}{\,\tilde{\rho}^{\,2}}\sin^2\theta\,
    \big[\,\tilde{a}\,\dd t-\big(\tilde{r}^2+(\tilde{a}+\tilde{l})^2\big)\,\dd\varphi\big]^2\Bigg],
    \label{GP-metric}
\end{align}
where, due to the factorization ${\tilde{\mathcal{P}}(\tilde{x})=(1-\tilde{x}^2)\tilde{P}(\tilde{x})}$ given by \eqref{tild_P_simpl},
\begin{align}
  \tilde{\Omega}       &= 1-\tilde{\alpha}\,\tilde{r}\,(\tilde{l}+\tilde{a} \cos\theta), \label{tilde-Omega-GP}\\
  \tilde{\rho}^{\,2}   &= \tilde{r}^{\,2}+(\tilde{l}+\tilde{a}\,\cos\theta)^2, \label{tilde-rho-GP}\\
  \tilde{P}(\theta)    &= a_0-a_3\cos\theta-a_4\cos^2\theta\,, \label{tilde-P-GPO}\\
  \tilde{Q}(\tilde{r}) &= (k'+e'^2 + g'^2) - 2m'\,\tilde{r} + \epsilon'\,\tilde{r}^{\,2}
        -2\tilde{\alpha}\,n'\,\tilde{r}^{\,3}
        -(\tilde{\alpha}^2 k' + c^2  \Lambda/3 )\,\tilde{r}^{\,4}\,. \label{tilde-Q-GP}
\end{align}
The PD parameters $k', e', g', m', \epsilon', n'$ are given by \eqref{direct_transformation_A-PD_parameters-simplified}--\eqref{coeff_Lambda}, while the acceleration ${\tilde{\alpha}=\alpha'}$ is expressed as \eqref{tild_alph}, i.e., \eqref{direct_transformation_A-PD-acceleration-explicit-again}. The constant $c$ is fixed by \eqref{C-definition} with \eqref{Cf-choice}, and the constants in \eqref{tilde-P-GPO} are explicitly determined by the relations
\begin{align}
a_0 &= \dfrac{k'}{\tilde{a}^2-\tilde{l}^2} - 2\tilde{\alpha}\,m'\, \tilde{l} + 3\lambda\,\tilde{l}^{2}\,, \label{a_0}\\
a_3 &= 2\tilde{\alpha}\,\tilde{a}\, m' - 4\lambda\,\tilde{a}\,\tilde{l}\,,\label{a_3}\\
a_4 &= - \lambda\,\tilde{a}^2\,, \label{a_4}
\end{align}
where ${\lambda = \tilde{\alpha}^{2}(k'+e'^{\,2}+g'^{\,2})+c^2 \Lambda/3}$, see \eqref{GP_k_rel-book}, \eqref{ai}, \eqref{Ai} and \eqref{definintion of lambda}.

In the original version of the GP metric \cite{GriffithsPodolsky:2005, GriffithsPodolsky:2006, PodolskyGriffiths:2006}, summarized in Eqs.~(16.18)--(16.20) in \cite{GriffithsPodolsky:2009}, the coefficient $a_0$ \emph{was set to one}, ${a_0=1}$. On the other hand, the twist parameter $\omega$ \emph{was kept undetermined}. Actually, they represent \emph{the same gauge freedom} encoded in rescaling of the coordinates and coefficients by specific powers of $a_0$ (see Eq.~\eqref{resc} below). We should also emphasize that the PD parameters ${k', e', g', m', \epsilon', n'}$ in the metric \eqref{GP-metric} --- and thus also the corresponding GP parameters ${\tilde{m}, \tilde{\alpha}, \tilde{a}, \tilde{l}, \tilde{e}, \tilde{g}}$ --- \emph{are dimensionless}. The coordinates $\tilde{r}, \theta$ involved in the metric \eqref{GP-metric} are also dimensionless, while $t, \varphi$ have dimension of length. However, it is easy to \emph{restore their proper physical dimension}. To this end --- similarly as in our previous paper \cite{OvcharenkoPodolskyAstorino:2024}, see Section~V.B  therein, --- we employ the rescaling by the parameter~$\gamma$ which has the dimension of length. It seems to be most convenient to choose
\begin{equation}
\gamma = c \,, \label{gamma=1/c}
\end{equation}
where ${c>0}$ is a unique constant defined in \eqref{C-definition}. Combining these two rescalings, namely
\begin{align}
    & \bar{t} = \sqrt{a_0}\,t\,, \qquad
      \bar{\varphi} = \frac{a_0}{c}\,\varphi  \,, \qquad
      \bar{r} = \frac{c}{\sqrt{a_0}}\,\tilde{r}\,, \nonumber\\
   &
      \bar{a} = \frac{c}{\sqrt{a_0}}\,\tilde{a}\,, \qquad
      \bar{l} = \frac{c}{\sqrt{a_0}}\,\tilde{l}\,, \qquad
      \bar{\alpha} = \frac{1}{c}\,\tilde{\alpha}\,,    \nonumber\\
    &
      \bar{k} = k'\,, \qquad\,\,
      \bar{\epsilon} = \frac{\epsilon'}{a_0}\,, \qquad
      \bar{n} = \frac{c}{a_0^{3/2}}\,n'\,,  \label{resc}\\
    &
      \bar{m} = \frac{c}{a_0^{3/2}}\,m'\,, \quad\,\,
      \bar{e} = \frac{c}{a_0}\,e'\,, \qquad
      \bar{g} = \frac{c}{a_0}\,g'\,, \nonumber\\
    &
      \bar{a}_0 = \frac{a_0}{c} \,, \qquad
      \bar{a}_3 = \frac{a_3}{a_0} \,, \qquad
      \bar{a}_4 = \frac{a_4}{a_0}  \,,  \nonumber
\end{align}
the metric \eqref{GP-metric} becomes
\begin{align}
    \dd s^2=\dfrac{1}{{\bar{\Omega}}^2}
    \Bigg[&-\dfrac{\bar{Q}}{\,\bar{\rho}^{\,2}}
    \Big[\dd \bar{t}-\big(\bar{a}\sin^2\theta+2\bar{l}(1-\cos\theta)\big)\dd\bar{\varphi}\Big]^2
     + \dfrac{\,\bar{\rho}^{\,2}}{\bar{Q}}\,\dd\bar{r}^2\nonumber\\
    &+ \dfrac{\,\bar{\rho}^{\,2}}{\bar{P}}\,\dd\theta^2
     + \dfrac{\bar{P}}{\,\bar{\rho}^{\,2}}\sin^2\theta\,
    \big[\,\bar{a}\,\dd \bar{t}-\big(\bar{r}^2+(\bar{a}+\bar{l})^2\big)\,\dd\bar{\varphi}\big]^2\Bigg],
    \label{GP-metric-final}
\end{align}
where
\begin{align}
  \bar{\Omega}     &= 1-\bar{\alpha}\,\bar{r}\,\bar{a}_0(\bar{l}+\bar{a} \cos\theta), \label{Omega-final}\\
  \bar{\rho}^{\,2} &= \bar{r}^{\,2}+(\bar{l}+\bar{a}\,\cos\theta)^2, \label{rho-final}\\
  \bar{P}(\theta)  &= 1-\bar{a}_3\cos\theta-\bar{a}_4\cos^2\theta\,, \label{P-final}\\
  \bar{Q}(\bar{r}) &= \Big(\dfrac{\bar{k}}{\,\bar{a}_0^2}+\bar{e}^{\,2}+\bar{g}^{\,2}\Big)
        -2\bar{m}\,\bar{r}+\bar{\epsilon}\,\bar{r}^{\,2}
        -2\bar{\alpha}\,\bar{a}_0\,\bar{n}\,\bar{r}^{\,3}
        -\Big(\bar{\alpha}^2 \bar{k} + \frac{\Lambda}{3} \Big)\,\bar{r}^{\,4}\,, \label{Q-final}
\end{align}
and
\begin{align}
 \bar{a}_3 &= 2\bar{\alpha}\,\bar{a}_0\,\bar{a}\, \bar{m}
- 4\bar{\alpha}^{2} \bar{a}_0^2\, \bar{a}\,\bar{l}\,\Big(\dfrac{\bar{k}}{\,\bar{a}_0^2}+\bar{e}^{\,2}+\bar{g}^{\,2}\Big)
  - 4\,\frac{\Lambda}{3}\,\bar{a}\,\bar{l}\,,\\
 \bar{a}_4 &= - \bar{\alpha}^{2} \bar{a}_0^2\,\bar{a}^2 \Big(\dfrac{\bar{k}}{\,\bar{a}_0^2}+\bar{e}^{\,2}+\bar{g}^{\,2}\Big)
 - \frac{\Lambda}{3}\,\bar{a}^2 \,.
\end{align}
With a straightforward identification
\begin{equation}
\omega \equiv \frac{1}{\bar{a}_0} = \frac{c}{a_0}\,, \label{omega=1/a0}
\end{equation}
this is exactly the GP metric given by Eqs.~(6.18)--(6.20) in the monograph~\cite{GriffithsPodolsky:2009}.
In particular, notice from \eqref{resc} that
\begin{align}
   &
      \bar{a} = \omega\,\sqrt{a_0}\,\tilde{a}\,, \qquad
      \bar{l} = \omega\,\sqrt{a_0}\,\tilde{l}\,.
\end{align}
where $\tilde{a}$ and $\tilde{l}$ are given by  \eqref{a^2-explicitly} and \eqref{l-explicitly}, respectively.
Moreover, with~\eqref{omega=1/a0} the formulae \eqref{C-definition} and  \eqref{a_0} give the \emph{convenient generic choice of the twist parameter} $\omega$. It is universal in the sense that it admits all subcases (in particular ${l=0, \alpha=0, a=0}$), without the unpleasant degeneracy of previous choices of $\omega$ which prevented to identify the subclass of accelerating NUT black holes in the GP coordinates. Note however that \emph{arbitrary} value of~$\omega$ can be restored by rescaling the acceleration parameter~$\tilde{\alpha}$ in the GP metric \eqref{GP-metric}, as described in Section~V.A of~\cite{OvcharenkoPodolskyAstorino:2024}.\\

Let us now discuss particular important subcases.

\newpage
\subsection{The special case ${\Lambda=0}$: no cosmological constant}
\label{sc:lambda=0GP}

For vanishing cosmological constant, the constants \eqref{c1},~\eqref{c2} with \eqref{direct_transformation_A-PD_parameters-simplified} read
\begin{align}
    c_1 & = -\big[1-\alpha^2(a^2+e^2+g^2-l^2)\big]^2+12\alpha^2(a^2+e^2+g^2-l^2-m^2) \,,\nonumber\\
    c_2 & = \big[1-\alpha^2(a^2+e^2+g^2-l^2)\big]^3   \\
        & \quad +36\alpha^2 \big[1-\alpha^2(a^2+e^2+g^2-l^2)\big] (a^2+e^2+g^2-l^2-m^2),\nonumber
\end{align}
thus
\begin{align}
\Big(c_2+\sqrt{c_1^3+c_2^2}\,\,\Big)^{1/3} &=
    +1-\alpha^2(a^2+e^2+g^2-l^2)+2\sqrt{3}\,\alpha\sqrt{a^2+e^2+g^2-l^2-m^2} \,,\nonumber\\
c_1 \Big(c_2+\sqrt{c_1^3+c_2^2}\,\,\Big)^{-1/3} &=
    -1+\alpha^2(a^2+e^2+g^2-l^2)+2\sqrt{3}\,\alpha\sqrt{a^2+e^2+g^2-l^2-m^2}\,,
\end{align}
and
\begin{align}
    2\epsilon' + c_1 \Big(c_2+\sqrt{c_1^3+c_2^2}\,\,\Big)^{-1/3} - \Big(c_2+\sqrt{c_1^3+c_2^2}\,\,\Big)^{1/3}
    = 12\big[1-\alpha^2(a^2-l^2)\big] \dfrac{l}{a}\dfrac{M}{I^2 \mp J^2}\,.
\end{align}
Now, using \eqref{direct_transformation_A-PD_parameters-simplified}
the coefficient $\kappa^2$, defined by \eqref{kappa} for ${j=0}$, becomes a \emph{full square}, so that
\begin{equation}
   \kappa = \frac{1}{I^2 \mp J^2}\, \Big[  \big[K-\alpha^2(a^2-l^2)\big]\,I\, M
          - \alpha^2\frac{l}{a}\big[1-\alpha^2(a^2-l^2)\big]
            \big[(a^2-l^2)\,L+ (e^2+g^2)\sqrt{I^2 \mp J^2} \,\big] \Big],\label{kappa-for-Lambda=0}
\end{equation}
and
\begin{equation}
   \lambda = \frac{I}{I^2 \mp J^2}\,\,
           \alpha^2 \Big[(a^2-l^2)\,L+ (e^2+g^2)\sqrt{I^2 \mp J^2}\,\Big]\,.\label{lambda-for-Lambda=0}
\end{equation}
Applying \eqref{l-explicitly} with a minus sign, and \eqref{a^2-explicitly}, also using the identity
\begin{equation}
   a^2\, (K-1) \big[K-\alpha^2(a^2-l^2)\big] = \alpha^2(a^2-l^2)^2\,,\label{identity}
\end{equation}
we obtain the explicit relation for the GP rotational parameters in a simple factorized form
\begin{align}
 \tilde{l}   &= \dfrac{1}{K-\alpha^2(a^2-l^2)}\,\dfrac{1-\alpha^2(a^2-l^2)}{1+\alpha^2(a^2-l^2)}\,\frac{l}{a}  \,, \label{l-explicitly-Lambda=0}\\
 \tilde{a}^2 &= \dfrac{1}{[K-\alpha^2(a^2-l^2)]^2}\,\dfrac{I^2 \mp J^2}{I^2}\,.
 \label{a^2-explicitly-Lambda=0}
\end{align}
These are exactly the expressions Eqs.~(74) and~(75) derived in our previous work \cite{OvcharenkoPodolskyAstorino:2024}, in which ${\Lambda=0}$ and ${\omega=1}$. More details, including also discussion of various subcases, are also given in \cite{OvcharenkoPodolskyAstorino:2024}. The results for vacuum black holes agree with \cite{WuWu:2024}.

\newpage
\subsection{The special case ${l=0}$: no NUT}
\label{sc:l=0GP}

In this case when the NUT parameter vanishes, using \eqref{defI-and -defJ-l=0}--\eqref{C-for-l=0} we obtain
\begin{align}
    c_1 & = -\big[1 - \alpha^2(a^2+e^2+g^2)\big]^{\,2} + 12 \Big[ \alpha^2 (a^2 + e^2 + g^2 - m^2) + \frac{\Lambda}{3} a^2 \Big]
         , \nonumber\\
    c_2 & =
    \big[1 - \alpha^2(a^2+e^2+g^2)\big]^3  + 18 \Lambda\,a^2\alpha^2 m^2    \label{c1c2-for-l=0}\\
        & \qquad
        + 36 \big[ 1 - \alpha^2(a^2+e^2+g^2)\big]\Big[ \alpha^2 (a^2 + e^2 + g^2 - m^2) + \frac{\Lambda}{3} a^2 \Big]
         . \nonumber
\end{align}
Therefore,
\begin{align}
    c_1^3+c_2^2 & =
         108\Big[ \alpha^2 (a^2 + e^2 + g^2 - m^2) + \frac{\Lambda}{3} a^2 \Big]
          \Big[\big[1 + \alpha^2(a^2+e^2+g^2)\big]^{\,2} + 4\, \Big( \frac{\Lambda}{3} a^2 - \alpha^2 m^2\Big)\Big]^2
         \nonumber\\
    &
    \qquad +36 \Lambda a^2\alpha^2 m^2  \Big(\big[1 - \alpha^2(a^2+e^2+g^2)\big]^3   \label{c1^3+c2^2-for-l=0}\\
    &
    \qquad\qquad + 36 \big[ 1 - \alpha^2(a^2+e^2+g^2)\big]\Big[ \alpha^2 (a^2 + e^2 + g^2 - m^2) + \frac{\Lambda}{3} a^2 \Big]
           + 9 \Lambda a^2\alpha^2 m^2 \Big), \nonumber
\end{align}
and
\begin{align}\label{kappa-for-l=0}
    \kappa^2 = \alpha^2m^2 - \frac{\lambda}{3}
    \Big[2 - 2\alpha^2(a^2+e^2+g^2)
    +c_1\Big(c_2+\sqrt{c_1^3+c_2^2}\,\,\Big)^{-1/3}-\Big(c_2+\sqrt{c_1^3+c_2^2}\,\,\Big)^{1/3}\,\Big],
\end{align}
where
\begin{equation}
\lambda = \alpha^2 (a^2 + e^2 + g^2) + \frac{\Lambda}{3}\,a^2  \,.
\end{equation}

Interestingly, the (complicated) expression \eqref{c1^3+c2^2-for-l=0} \emph{considerably simplifies whenever}
\begin{align}
  \Lambda\,a\,\alpha\,m = 0\,,
\end{align}
i.e., if  ${\Lambda=0}$ or ${a=0}$ or ${\alpha=0}$ or ${m=0}$.  In such a case,
\begin{align}
    \sqrt{c_1^3+c_2^2} & =
         6\sqrt{3}\,\sqrt{\lambda-\alpha^2m^2}\,
          \Big[\big[1 - \alpha^2(a^2+e^2+g^2)\big]^{\,2} + 4\,(\lambda-\alpha^2 m^2 )\Big],
\end{align}
implying
\begin{align}
\Big(c_2+\sqrt{c_1^3+c_2^2}\,\,\Big)^{1/3} &=
    +1-\alpha^2(a^2+e^2+g^2)+2\sqrt{3}\,\sqrt{\lambda-\alpha^2 m^2} \,,\nonumber\\
c_1 \Big(c_2+\sqrt{c_1^3+c_2^2}\,\,\Big)^{-1/3} &=
    -1+\alpha^2(a^2+e^2+g^2)+2\sqrt{3}\,\sqrt{\lambda-\alpha^2 m^2}\,,
\end{align}
and thus \eqref{kappa-for-l=0} reduces to ${\kappa=\alpha\,m}$. Since ${\tilde{\alpha}\,m'=\alpha'm'=\alpha\,m}$, see \eqref{acceleration}, \eqref{direct_transformation_A-PD_parameters-l=0}, by taking the minus sign in \eqref{l-explicitly} we obtain
\begin{align}\label{l=0-implies-tildel=0}
    \tilde{l}=0\,.
\end{align}
In view of \eqref{resc}, the corresponding NUT parameter $\bar{l}$ with proper physical dimension also vanishes,
\begin{align}\label{l-and-a-explicitly l=0}
    \bar{l}=0   \,.
\end{align}

However, to properly evaluate the complementary Kerr-like parameter $\bar{a}$, one has to perform careful limits in the general case when also
\begin{align}
  \Lambda\,a\,\alpha\,m \ne 0\,.
\end{align}
We will proceed by separately considering the distinct subcases, and evaluate the corresponding limits of \eqref{c1c2-for-l=0}.

\subsubsection*{$\bullet$ The ${\Lambda=0}$ subcase}

\noindent
Expanding \eqref{kappa-for-l=0} in $\Lambda$, the leading term of $\kappa^2$ is
    \begin{align}
        \kappa^2=\alpha^2m^2+O(\Lambda)\,.
    \end{align}
Taking the minus sign in expression \eqref{l-explicitly} for $\tilde{l}$, and then using \eqref{a^2-explicitly}, the limit ${\Lambda\to 0}$ reads
\begin{align}
    \tilde{l}&=-\dfrac{\Lambda}{3}\dfrac{\alpha\,m\,a^2}{[1+\alpha^2(a^2+e^2+g^2)]^2-4\alpha^2 m^2}+O(\Lambda^2)\to 0\,, \nonumber\\
    \tilde{a}&= 1+O(\Lambda) \to  1\,. \label{tilde-a-for-l=0,Lambda=0}
\end{align}
Moreover, in the present case ${l=0}$ (and thus ${\tilde{l}=0}$)  there is
\begin{equation}\label{c-and-a_0-foir-l=0}
c = a\,, \qquad  a_0 = 1
\end{equation}
(see \eqref{C-for-l=0} and \eqref{a_0} where ${k'=1}$), so that from \eqref{resc}
we obtain the NUT parameter ${\bar{l}=a\,\tilde{l}}$ and the Kerr parameter ${\bar{a}=a\,\tilde{a}}$ with proper physical dimension of length, namely
\begin{align}\label{l-and-a-explicitly l=0 Lambda=0}
    \bar{l}=0\,,  \qquad \bar{a}=a\,.
\end{align}
There is thus an agreement between the Griffiths-Podolsk\'y and the Astorino parameters in this subcase ${l=0}$ with ${\Lambda=0}$.

\newpage

\subsubsection*{$\bullet$ The ${a=0}$ subcase}

\noindent
The leading term of $\kappa^2$ obtained by --- expanding in terms of $a^2$ --- the expression \eqref{kappa-for-l=0} with a general \eqref{c1c2-for-l=0} is
    \begin{align}
        \kappa^2=\alpha^2m^2\Big(1+\dfrac{\Lambda}{3}\dfrac{4\alpha^2a^2(e^2+g^2)}{[1+\alpha^2(e^2+g^2)]^2-4\alpha^2m^2}\Big)+O(a^4)\,.
    \end{align}
Considering the minus sign in \eqref{l-explicitly}, and using \eqref{a^2-explicitly}, for ${a \to 0}$ we obtain
    \begin{align}
        \tilde{l}&=-\dfrac{\alpha\, m\, \Lambda a^2}{[1+\alpha^2(e^2+g^2)]^2-4\alpha^2m^2}\rightarrow 0 \,, \nonumber\\
        \tilde{a}&=1+O(a^2)\rightarrow 1 \,,
    \end{align}
as in \eqref{tilde-a-for-l=0,Lambda=0}. From \eqref{resc} with the parameters \eqref{c-and-a_0-foir-l=0}, that is ${c = a}$, ${a_0 = 1}$, we thus get
\begin{align}\label{l-and-a-explicitly l=0 a=0}
    \bar{l}=0\,,  \qquad \bar{a}=0\,.
\end{align}
This is the same as \eqref{l-and-a-explicitly l=0 Lambda=0}, but now for ${a=0}$.

We can also explicitly derive the metric functions of the GP metric \eqref{GP-metric-final}.
Recall the PD parameters \eqref{direct_transformation_A-PD_parameters-l=0} for the ${l=0}$ case, namely
\begin{align}\label{PD_parameters_l=0}
\alpha' &= \tilde{\alpha} = \alpha\,a \,, \qquad
k' = 1 \,, \qquad
n' = -\alpha\,m \,, \qquad
\alpha' m' = \alpha\,m \,, \nonumber\\
\epsilon'  &= 1 - \alpha^2(a^2+e^2+g^2)\,, \qquad
({e'}^2 + {g'}^2) = ({e}^2 + {g}^2)/a^{2}\,.
\end{align}
Therefore, $\lambda$ given by \eqref{definintion of lambda} for ${a \to 0}$ becomes
\begin{equation}
\lambda = \alpha^2(e^2+g^2) \,.
\end{equation}
Applying  \eqref{a_3}, \eqref{a_4} for ${\tilde{l}=0}$ and ${\tilde{a}=1}$ we evaluate the coefficients
    \begin{align}
       a_3 &=2\tilde{\alpha}\,\tilde{a}\,m'=2\alpha m\,, \\
       a_4 &=-\lambda \tilde{a}^2=-\alpha^2(e^2+g^2)\,.
    \end{align}
The physical GP parameters are then obtained by the rescaling  \eqref{resc} with \eqref{c-and-a_0-foir-l=0},
\begin{align}
   &
      \bar{a} = a\,\tilde{a} \,, \qquad
      \bar{l} = a\,\tilde{l} \,, \qquad
      \bar{\alpha} = \frac{1}{a}\,\tilde{\alpha} = \alpha \,,    \nonumber\\
    &
      \bar{k} = k' = 1\,, \qquad
      \bar{\epsilon} = \epsilon' = 1 - \alpha^2(e^2+g^2)\,, \qquad
      \bar{n} = a\,n' = -\alpha\,a\,m\,,  \label{resc-a=0}\\
    &
      \bar{m} = a\,m' = m\,, \qquad
      \bar{e} = a\,e' = e\,, \qquad
      \bar{g} = a\,g' = g\,, \nonumber\\
    &
      \bar{a}_0 = \frac{1}{a} \,, \qquad
      \bar{a}_3 = a_3 = 2\alpha m\,, \qquad
      \bar{a}_4 = a_4 = -\alpha^2(e^2+g^2)\,,  \nonumber
\end{align}
so that ${\bar{a}_0\,\bar{a} = \tilde{a} \to 1 }$ and
${\bar{a}_0\,\bar{l} = \tilde{l} \to 0 }$. The GP metric functions of \eqref{GP-metric-final} thus become
\begin{align}
  \bar{\Omega}     &= 1-\alpha\,\bar{r} \cos\theta\,, \label{Omega_l=0,a=0}\\
  \bar{\rho}^{\,2} &= \bar{r}^{\,2}\,, \label{rho_l=0,a=0}\\
  \bar{P}  &= 1-2\alpha m\,\cos\theta+\alpha^2(e^2+g^2)\,\cos^2\theta\,, \label{P_l=0,a=0}\\
  \bar{Q} &= e^2+g^2
        -2m\,\bar{r}+[1 - \alpha^2(e^2+g^2)]\,\bar{r}^{\,2}
        +2\alpha^2 m\,\bar{r}^{\,3}
        -\Big(\alpha^2 + \frac{\Lambda}{3} \Big)\,\bar{r}^{\,4}\nonumber\\
        &= (\bar{r}^{\,2}-2m\,\bar{r}+e^2+g^2)(1-\alpha^2 \bar{r}^{\,2})-\frac{\Lambda}{3}\,\bar{r}^{\,4}\,, \label{Q_l=0,a=0}
\end{align}
and the metric is explicitly
\begin{align}
    \dd s^2=\dfrac{1}{{\bar{\Omega}}^2}
    \Bigg[&-\dfrac{\bar{Q}}{\,\bar{r}^{\,2}}\, \dd \bar{t}^{\,2}
     + \dfrac{\,\bar{r}^{\,2}}{\bar{Q}}\,\dd\bar{r}^2
     + \dfrac{\,\bar{r}^{\,2}}{\bar{P}}\,\dd\theta^2
     + \bar{r}^{\,2} \bar{P}\sin^2\theta\,\dd\bar{\varphi}^{\,2}\Bigg].
    \label{GP-metric_l=0,a=0}
\end{align}
This result fully corresponds to the expressions for accelerating charged black holes with a cosmological constant (the charged C-metric with $\Lambda$, when ${l=0=a}$), as given by Eqs.~(14.43) and ~(14.45) in the monograph \cite{GriffithsPodolsky:2009}, and also by Eqs.~(71)--(78) in \cite{PodolskyVratny:2023}.

\subsubsection*{$\bullet$ The ${\alpha=0}$ subcase}

\noindent
Expanding \eqref{kappa-for-l=0} with \eqref{c1c2-for-l=0} in $\alpha^2$, the dominant term is
    \begin{align}
        \kappa^2=\alpha^2m^2 \,\dfrac{(1+\tfrac{2}{3}\Lambda a^2)^2}
        {1+\tfrac{4}{3}\Lambda a^2}+O(\alpha^4)\,.
    \end{align}
Then \eqref{l-explicitly}, with the minus sign, and \eqref{a^2-explicitly} for ${\alpha \to 0}$ yield
    \begin{align}
\tilde{l}&=\alpha m\,\bigg(1-\dfrac{1+\tfrac{2}{3}\Lambda a^2}{\sqrt{1+\tfrac{4}{3}\Lambda a^2}}\bigg)  \dfrac{1}{\tfrac{2}{3}\Lambda a^2} +O(\alpha^3)\to 0\,, \nonumber\\
\tilde{a}^2&=\bigg(\sqrt{1+\tfrac{4}{3}\Lambda a^2}-1\bigg)\dfrac{1}{\tfrac{2}{3}\Lambda a^2} +O(\alpha^2)\,. \label{tilde-l=0-alpha=0}
    \end{align}
From \eqref{a_0}, using ${k'=1}$, we thus obtain
\begin{align}
a_0 &= \dfrac{1}{\tilde{a}^2} \,,
\end{align}
and with ${c=a}$ the relations \eqref{resc} for ${\alpha\to0}$ imply
 \begin{align}
   \bar{l}&= a\,\tilde{a}\,\tilde{l} = 0 \,, \nonumber\\
   \bar{a}&= a\,\tilde{a}^2 = \dfrac{\sqrt{1+\tfrac{4}{3}\Lambda a^2}-1}{\tfrac{2}{3}\Lambda a}\,,\label{l=0-alpha=0}
 \end{align}
which can be inverted as
 \begin{align}
   a = \dfrac{\bar{a}}{1-\tfrac{1}{3}\Lambda \bar{a}^{\,2}}\,.\label{a-for-l=0-alpha=0}
 \end{align}
Clearly, in this case ${\bar{a}\ne a}$. But \emph{for small values of} ${\Lambda a^2}$ (and small values of $\Lambda \bar{a}^{\,2}$)  we get
 \begin{align}
   \bar{a} \approx a\,,
 \end{align}
which means that in the limit ${\Lambda a^2 \to 0}$ the GP Kerr-like rotation parameter $\bar{a}$ agrees with the A$^+$  Kerr-like parameter $a$.

The GP metric functions are obtained from the PD parameters \eqref{direct_transformation_A-PD_parameters-l=0} for the ${l=0}$ case with ${\alpha=0}$, that is\begin{align}\label{PD_parameters_l=0,alpha=0}
\alpha' &= \tilde{\alpha} = \alpha\,a =0  \,, \qquad
k' = 1 \,, \qquad
n' = 0 \,, \qquad
\alpha' m' = \alpha\,m = 0\,, \nonumber\\
\epsilon'  &= 1 \,, \qquad
({e'}^2 + {g'}^2) = ({e}^2 + {g}^2)/a^{2}\,.
\end{align}
Moreover, $\lambda$ given by \eqref{definintion of lambda} now reduces to
\begin{equation}
\lambda = \tfrac{1}{3}\Lambda a^2 \,,
\end{equation}
and using  \eqref{a_0}--\eqref{a_4} for ${\tilde{l}=0}$ and ${\tilde{\alpha}=0}$ we obtain
    \begin{align}
    a_0 &= \dfrac{1}{\tilde{a}^2} = \dfrac{a}{\bar{a}} =  \dfrac{1}{1-\tfrac{1}{3}\Lambda \bar{a}^{\,2}} \,, \label{a_0next}\\
    a_3 &= 0\,, \\
    a_4 &=-\lambda \tilde{a}^2 = - \dfrac{\tfrac{1}{3}\Lambda\bar{a}^2}{1-\tfrac{1}{3}\Lambda \bar{a}^{\,2}} \,.
    \end{align}

The physical parameters of the GP metric are obtained by the rescaling  \eqref{resc} of \eqref{PD_parameters_l=0,alpha=0} with ${c=a}$ and \eqref{a_0next},
\begin{align}
   &
      \bar{\alpha} = \frac{1}{a}\,\tilde{\alpha} = \alpha = 0 \,,\qquad
      \bar{k} = k' = 1\,, \qquad
      \bar{n} = 0\,,    \nonumber\\
    &
      \bar{\epsilon} = \frac{\epsilon'}{a_0} = 1-\tfrac{1}{3}\Lambda \bar{a}^{\,2}\,,\qquad
      \bar{m} = \frac{a}{a_0^{3/2}}\,m'
      = \frac{a}{a_0^{3/2}}\,\frac{m}{a}
      = (1-\tfrac{1}{3}\Lambda \bar{a}^{\,2})^{3/2}\,m \,,    \label{resc-l=0,alpha=0}\\
    &
      \bar{e} = \frac{a}{a_0}\,e' = \bar{a}\,\frac{e}{a} = (1-\tfrac{1}{3}\Lambda \bar{a}^{\,2})\,e\,, \qquad
      \bar{g} = \frac{a}{a_0}\,g' = \bar{a}\,\frac{g}{a} = (1-\tfrac{1}{3}\Lambda \bar{a}^{\,2})\,g\,, \nonumber\\
    &
      \bar{a}_0 = \frac{a_0}{a} = \frac{1}{\bar{a}}\,, \qquad
      \bar{a}_3 = \frac{a_3}{a_0} = 0\,, \qquad
      \bar{a}_4 = \frac{a_4}{a_0} = -\tfrac{1}{3}\Lambda\bar{a}^2 \,,  \nonumber
\end{align}
The metric functions \eqref{Omega-final}--\eqref{Q-final} of the GP metric \eqref{GP-metric-final} thus become
\begin{align}
  \bar{\Omega}     &= 1\,, \label{Omega_l=0,alpha=0}\\
  \bar{\rho}^{\,2} &= \bar{r}^{\,2}+\bar{a}^{\,2}\cos^2\theta\,, \label{rho_l=0,alpha=0}\\
  \bar{P}(\theta)  &= 1+\tfrac{1}{3}\Lambda\bar{a}^2\cos^2\theta\,, \label{P_l=0,alpha=0}\\
  \bar{Q}(\bar{r}) &= (\bar{a}^{\,2}+\bar{e}^{\,2}+\bar{g}^{\,2})
        -2\bar{m}\,\bar{r} + \bar{r}^{\,2}
        -\frac{\Lambda}{3}\,\bar{r}^{\,2}\,( \bar{a}^{\,2} + \bar{r}^{\,2})\,. \label{Q_l=0,alpha=0}
\end{align}
This is exactly the metric of Kerr-Newman-(anti-)de~Sitter black hole, as presented e.g. in \cite{GriffithsPodolsky:2009} in  Eq.~(16.23) with ${l=0}$, and in~(16.24) with ${\alpha=0}$ (see also \cite{PodolskyVratny:2023}, namely Eqs.~(59)--(62) and Eqs.~(77)--(78), respectively). However, we should emphasize, that the GP and A parameters are \emph{different}, namely
 \begin{align}
   \bar{a}&= \dfrac{\sqrt{1+\tfrac{4}{3}\Lambda a^2}-1}{\tfrac{2}{3}\Lambda a}\,, \qquad
       \bar{m} =  \sqrt{(1-\tfrac{1}{3}\Lambda \bar{a}^{\,2})^3}\,\,m \,, \\
   \bar{e}& = (1-\tfrac{1}{3}\Lambda \bar{a}^{\,2})\,e\,, \qquad\qquad
      \bar{g} = (1-\tfrac{1}{3}\Lambda \bar{a}^{\,2})\,g\,.
\end{align}
Such a difference is not surprising because the A$^+$ metric functions \eqref{delta_r_init}, \eqref{delta_x_init}
in this ${\alpha=0}$ subcase represent an alternative form of the Kerr-Newman-(anti-)de~Sitter black hole. To obtain the form given in \cite{GriffithsPodolsky:2009}, an additional reparametrization is required.\\

To summarize, in all the three subcases ${\Lambda=0}$, ${a=0}$, ${\alpha=0}$ we thus obtained that ${l=0}$ implies ${\bar{l}=0}$, but ${ \bar{a} \ne a}$ when ${\Lambda a^2 \ne 0}$.

\subsection{The special case ${\alpha=0}$: no acceleration}
\label{sc:alpha=0GP}

In case of vanishing acceleration the constants \eqref{c1},~\eqref{c2} with \eqref{direct_transformation_A-PD_parameters-alpha=0} are simply
\begin{align}\label{c2,c3-for-alpha=0}
    c_1=4\Lambda a^2-(1+\Lambda l^2)^2\,,    \qquad
    c_2=(1+\Lambda l^2)^3+12\Lambda a^2(1-\Lambda l^2)\,,
\end{align}
so that
\begin{align}\label{c2+c3-for-alpha=0}
c_1^3+c_2^2 = 4\Lambda a^2 \big[\, 3(3-\Lambda l^2)(1+\Lambda l^2)^3
  + 8\Lambda a^2 (3 + 2\Lambda a^2 - 12 \Lambda l^2 + 3 (\Lambda l^2)^2\, \big]\,,
\end{align}
and
\begin{align}\label{kappa-for-alpha=0}
    \kappa^2 = - \frac{\lambda}{3}
    \Big[2(1 - \Lambda l^2)
    +c_1\Big(c_2+\sqrt{c_1^3+c_2^2}\,\,\Big)^{-1/3}-\Big(c_2+\sqrt{c_1^3+c_2^2}\,\,\Big)^{1/3}\,\Big].
\end{align}
It is also useful now to recall the PD parameters \eqref{direct_transformation_A-PD_parameters-alpha=0}, \eqref{C-for-alpha=0} for the ${\alpha=0}$ case, namely
\begin{align}\label{PD_parameters_alpha=0}
\alpha' &= \tilde{\alpha} = \alpha\,a = 0 \,, \qquad
k' = 1 - \frac{l^2}{a^2} \,, \qquad
n' = \frac{l}{a}\Big( 1 -\frac{\Lambda}{3}\,l^2 \Big) \,, \qquad
m' = m/a \,, \nonumber\\
\epsilon'  &= 1 - \Lambda\,l^2\,, \qquad
({e'}^2 + {g'}^2) = ({e}^2 + {g}^2)/a^{2}\,, \qquad
c = a  \,.
\end{align}
so that
\begin{equation}\label{lambda-alpha=0}
\lambda =  \tfrac{1}{3} \Lambda a^2  \,.
\end{equation}

General expressions for $\tilde{l}$ and $\tilde{a}$, given by \eqref{l-explicitly} and \eqref{a^2-explicitly},
are quite complicated. In order to get further insight and achieve simplifications, we will assume the subcases when some of the other parameters tend zero.

\subsubsection*{$\bullet$ The ${\Lambda=0}$ subcase}

\noindent
Expanding \eqref{kappa-for-alpha=0} in $\Lambda$, the leading term is
    \begin{align}
        \kappa = \tfrac{2}{3}\, a l\, \Lambda + O(\Lambda^2)\,.
    \end{align}
Taking the plus sign in expression \eqref{l-explicitly} for $\tilde{l}$ and  using \eqref{a^2-explicitly}, the limit ${\Lambda\to 0}$ gives
\begin{align}
    \tilde{l} = \dfrac{l}{a} +O(\Lambda) \to  \dfrac{l}{a}\,,  \qquad
    \tilde{a} = 1+O(\Lambda) \,\,\to  1\,.  \label{tilde-a-for-alpha=0,Lambda=0}
\end{align}
By applying \eqref{a_0} with \eqref{PD_parameters_alpha=0} we obtain ${a_0=1}$, and then \eqref{resc} with ${c=a}$ give the Kerr-like and NUT parameters with proper dimensionality as
\begin{align}
    \bar{l}=l\,,\qquad \bar{a}=a\,.
\end{align}
The GP and A$^+$ physical rotational parameters thus coincide when ${\alpha=0=\Lambda}$.

\subsubsection*{$\bullet$ The ${l=0}$ subcase}

\noindent
Expansion of \eqref{kappa-for-alpha=0} in $l^2$ gives  the dominant term
    \begin{align}
        \kappa^2 = \dfrac{\tfrac{4}{9}\Lambda^2a^2\,l^2}{1+\tfrac{4}{3}\Lambda a^2} +O(l^4)\,.
    \end{align}
Then \eqref{l-explicitly}, with the plus sign, and \eqref{a^2-explicitly} in the limit ${l \to 0}$ yields
    \begin{align}
\tilde{l}  &= \dfrac{1}{\sqrt{1+\tfrac{4}{3}\Lambda a^2}}\,  \dfrac{l}{a} +O(l^3)\to 0\,, \nonumber\\
\tilde{a}^2&= \bigg(\sqrt{1+\tfrac{4}{3}\Lambda a^2}-1\bigg)\dfrac{1}{\tfrac{2}{3}\Lambda a^2} +O(l^2) \to \dfrac{\sqrt{1+\tfrac{4}{3}\Lambda a^2}-1}{\tfrac{2}{3}\Lambda a^2}\,.
    \end{align}
We have thus obtained the same expressions \eqref{tilde-l=0-alpha=0} as in the ${\alpha=0}$ subcase of the case ${l=0}$, and therefore the same result as in \eqref{l=0-alpha=0}, that is
 \begin{align}
   \bar{l} =  0 \,, \qquad
   \bar{a} =  \dfrac{\sqrt{1+\tfrac{4}{3}\Lambda a^2}-1}{\tfrac{2}{3}\Lambda a}\,.\label{alpha=0-l=0}
 \end{align}
This demonstrates that the order in which the limits ${l \to 0}$ and ${\alpha \to 0}$ are performed is irrelevant, so that the procedure of obtaining these subcases from the general Griffiths-Podolsk\'y metric form commutes.

\subsubsection*{$\bullet$ The ${a=0}$ subcase}

\noindent
The dominant term in the expansion of $\kappa^2$ is now
\begin{align}
    \kappa^2 = \tfrac{4}{9}\Lambda^2a^2\,l^2 + O(a^4)\,.
\end{align}
Taking the plus sign in \eqref{l-explicitly} and \eqref{a^2-explicitly}, in the limit ${a \to 0}$ we obtain
    \begin{align}
\tilde{l}  &= \dfrac{l}{a}+O(a)\,, \nonumber\\
\tilde{a}^2&= \dfrac{1}{1+\Lambda l^2}+O(a^2)\,.
    \end{align}
Using \eqref{PD_parameters_alpha=0} and \eqref{lambda-alpha=0}, from \eqref{a_0} in the limit ${a \to 0}$ we get
\begin{align}
    a_0=1+\Lambda l^2\,.
\end{align}
The scaling relations \eqref{resc} for ${a\to0}$ then imply
 \begin{align}
   \bar{l} =  \dfrac{l}{\sqrt{1+\Lambda l^2}} \,, \qquad
   \bar{a} =  0\,.\label{alpha=0-a=0}
 \end{align}
In general, ${\bar{l}\ne l}$. The inverse relation is
\begin{align}
   l = \dfrac{\bar{l}}{\sqrt{1-\Lambda \bar{l}^{\,2}}}\,.\label{l-for-a=0-alpha=0}
 \end{align}
It can be immediately seen that \emph{for small values of} ${\Lambda l^2}$ (and small values of $\Lambda \bar{l}^{\,2}$)  we get
 \begin{align}
   \bar{l} \approx l\,.
 \end{align}

From \eqref{a_0}--\eqref{a_4} with \eqref{PD_parameters_alpha=0}, \eqref{lambda-alpha=0}, for ${\alpha=0}$ and ${a\to0}$ we also obtain
    \begin{align}\label{a_0next-l}
    a_0 = \dfrac{1}{1-\Lambda \bar{l}^{\,2}} \,, \qquad
    a_3 = 0\,, \qquad
    a_4 = 0 \,.
    \end{align}
The GP metric functions are now obtained from the dimensionless PD parameters
\eqref{PD_parameters_alpha=0} using the rescaling \eqref{resc},
\begin{align}
   &
      \bar{\alpha} = \frac{1}{a}\,\tilde{\alpha} = \alpha = 0 \,,\qquad
      \bar{k} = k' = 1 - \frac{l^2}{a^2}\,, \qquad
      \bar{\epsilon} = \frac{\epsilon'}{a_0} = \frac{1 - \Lambda l^2}{1+\Lambda l^2}
      =1 - 2\Lambda\bar{l}^{\,2} \,,   \nonumber\\
    &
      \bar{n} = \frac{a}{a_0^{3/2}}\,\,n'
      = l\, \frac{ 1 - \tfrac{1}{3} \Lambda l^2 }{(1+\Lambda l^2)^{3/2}}\,,\qquad
      \bar{m} = \frac{a}{a_0^{3/2}}\,m'
      = \frac{m}{(1+\Lambda l^2)^{3/2}} \,,    \label{resc-l=0,a=0}\\
    &
      \bar{e} = \frac{a}{a_0}\,e' = \frac{e}{1+\Lambda l^2} \,, \qquad
      \bar{g} = \frac{a}{a_0}\,g' = \frac{g}{1+\Lambda l^2}\,, \nonumber\\
    &
      \bar{a}_0 = \frac{a_0}{a} = \frac{1+\Lambda l^2}{a}\,, \qquad
      \bar{a}_3 = \frac{a_3}{a_0} = 0\,, \qquad
      \bar{a}_4 = \frac{a_4}{a_0} = 0 \,.  \nonumber
\end{align}
The metric functions \eqref{Omega-final}--\eqref{Q-final} of the GP metric \eqref{GP-metric-final} thus become
\begin{align}
  \bar{\Omega}     &= 1\,, \label{Omega_alpha=0,a=0}\\
  \bar{\rho}^{\,2} &= \bar{r}^{\,2} + \bar{l}^{\,2}\,, \label{rho_alpha=0,a=0}\\
  \bar{P}  &= 1  \,, \label{P_alpha=0,a=0}\\
  \bar{Q} &= \bar{e}^{\,2}+\bar{g}^{\,2} - \frac{l^2}{(1+\Lambda l^2)^2}
        -2\bar{m}\,\bar{r} + (1 - 2\Lambda\bar{l}^{\,2})\,\bar{r}^{\,2}
        -\frac{\Lambda}{3}\,\bar{r}^{\,4}\, \label{Q_alpha=0,a=0}
\end{align}
which, applying the identity
   \begin{align}
        \frac{l^2}{(1+\Lambda l^2)^2}= (1-\Lambda\bar{l}^{\,2})\,\bar{l}^{\,2}\,,
    \end{align}
can be expressed as
\begin{align}
  \bar{Q} = \bar{r}^{\,2}-2\bar{m}\,\bar{r} + \bar{e}^{\,2}+\bar{g}^{\,2} - \bar{l}^{\,2}
    - \Lambda \big(\! -\bar{l}^{\,4} + 2\bar{l}^{\,2} \bar{r}^{\,2} + \tfrac{1}{3}\bar{r}^{\,4} \big) \,. \label{Q_alpha=0,a=0final}
\end{align}
This fully agrees with the metric of charged NUT--(anti-)de~Sitter black hole presented in \cite{GriffithsPodolsky:2009} in  Eq.~(16.23) for ${a=0}$ (see also Eqs.~(69), (70) in \cite{PodolskyVratny:2023}).

Finally, we emphasize the difference between the GP and A physical parameters, namely
 \begin{align}
   \bar{l}&= \dfrac{l}{\sqrt{1+\Lambda l^2}}\,, \qquad
       \bar{m} =  \frac{m}{\sqrt{(1+\Lambda l^2)^3}} \,, \\
   \bar{e}& = \frac{e}{1+\Lambda l^2}\,, \qquad\qquad
      \bar{g} = \frac{g}{1+\Lambda l^2}\,.
\end{align}
They agree only in the limit ${\Lambda l^2 \to 0}$.

\subsection{The special case ${a=0}$: no Kerr-like rotation}
\label{sc:a=0GP}

In this case, for which
\begin{align}\label{direct_transformation_A-PD_parameters-a=0-again}
\alpha' &= \tilde{\alpha} = \alpha^2 l^2 \,, \nonumber\\
k'  &= \frac{\alpha^2 l^2-1}{4\alpha^2 l^4}\,\big(2m\,l +e^2+g^2   \big) - \dfrac{\Lambda}{3\alpha^2} \,, \nonumber\\
n'  &= \frac{(m-l)\,l+(e^2+g^2)}{2 \alpha\, l^3} + \frac{\alpha}{2}(m+l) + \dfrac{\Lambda l}{6 \alpha}\,,\\
\epsilon'  &= -2 (1+\alpha^2l^2) + \frac{1}{2}(e^2+g^2)\Big( \alpha^2 + \frac{3}{l^2}  \Big)  , \nonumber\\
\alpha' m' &= \frac{\alpha}{2}\,\Big[ -(m+l) + \alpha^2l^2(l-m)
    +\frac{e^2+g^2}{l} + \dfrac{\Lambda}{3}\,l^3 \,\Big] ,\nonumber\\
{e'}^2 + {g'}^2 &= \frac{1-\alpha^2l^2}{2\alpha^2 l^4}\,({e}^2 + {g}^2)\,, \nonumber\\
 c^2 &= 2\alpha^2\l^4 \,, \nonumber
\end{align}
see \eqref{direct_transformation_A-PD_parameters-a=0} and \eqref{C-for-a=0}, the key constants \eqref{c1},~\eqref{c2} are much more involved, namely
\begin{align}\label{c2,c3-for-a=0}
    c_1=& -\big[1-\alpha^2(e^2+g^2)\big]^2+12\alpha^2 (e^2+g^2-m^2)\nonumber\\
    & - l^2\big[(\alpha^2-\Lambda)^2l^2-2\alpha^2(\alpha^2+\Lambda)(e^2+g^2)+14\alpha^2+2\Lambda\big],\\
    c_2=& -1 + \big[1-\alpha^2(e^2+g^2)\big]^3 +(1+\Lambda l^2)^3
          +36\alpha^2\big[1-\alpha^2(e^2+g^2)\big](e^2+g^2-m^2)
    \nonumber\\
    & +3\alpha^2l^2\Big[(1+2l^2\Lambda)^2+6\Lambda (m^2-l^2+e^2+g^2)-12+\Lambda(e^2+g^2-l^2)(4-\Lambda l^2)\Big] \nonumber\\
    & +3\alpha^4l^2 \Big[\, l^2 + 12(e^2+g^2-m^2-l^2)+6\Lambda l^2(l^2+m^2-e^2-g^2) \nonumber\\
    & \hspace{16mm} \Lambda \big[(e^2+g^2)^2-l^4\big]+10(e^2+g^2)\Big] \nonumber\\
    & +\alpha^6\big[(l^2-e^2-g^2)^3+(e^2+g^2)^3\big].
\end{align}
To proceed further, let us consider the specific subcases.

\subsubsection*{$\bullet$ The ${\Lambda=0}$ subcase}

\noindent
Expanding \eqref{kappa} with \eqref{direct_transformation_A-PD_parameters-a=0-again}--\eqref{c2,c3-for-a=0} in powers of $\Lambda$, the leading term is
    \begin{align}
        \kappa = \tfrac{1}{2}\,\alpha \big[ \, m-l +\alpha^2[(m+l)l -e^2-g^2] \big] + O(\Lambda)\,.
    \end{align}
From \eqref{l-explicitly} and \eqref{a^2-explicitly} we then obtain the  ${\Lambda\to 0}$ limit as
\begin{align}
    \tilde{l} = \dfrac{1}{\alpha l}\,\dfrac{1+\alpha^2 l^2}{1-\alpha^2 l^2}\,,  \qquad
    \tilde{a} = \dfrac{2}{1-\alpha^2 l^2}\,.  \label{tilde-a-for-a=0,Lambda=0}
\end{align}
These are exactly the expressions (124) in our previous work \cite{OvcharenkoPodolskyAstorino:2024} (for the choice ${\omega_0=1}$).

Physical parameters of the GP metric are then  obtained by the rescaling \eqref{resc}. Because
${(\tilde{a}^2-\tilde{l}^2)^{-1} = -\alpha^2l^2}$ and ${\lambda =  \tfrac{1}{4}\alpha^2(1-\alpha^2l^2)\big(e^2+g^2-2ml)}$, from \eqref{a_0} and \eqref{direct_transformation_A-PD_parameters-a=0-again} we obtain
\begin{align} \label{a_0-for-a=0-Lambda=0}
a_0 = \dfrac{1-2\alpha^2ml+\alpha^4l^2(e^2+g^2-l^2)}{1-\alpha^2 l^2}\,.
\end{align}
Therefore, with ${c=\sqrt2\,\alpha\,l^2}$,
\begin{align} \label{a=0-Lambda=0}
      \bar{\alpha} &= \frac{1}{c}\,\tilde{\alpha}= \frac{\alpha}{\sqrt2} \,,\nonumber\\
      \bar{a} &= \frac{c}{\sqrt{a_0}}\,\tilde{a} = \sqrt{\frac{2}{a_0}}\,\dfrac{2\,\alpha\,l^2}{1-\alpha^2 l^2}  \,, \\
      \bar{l} &= \frac{c}{\sqrt{a_0}}\,\tilde{l} = \sqrt{\frac{2}{a_0}}\,\dfrac{1+\alpha^2 l^2}{1-\alpha^2 l^2}\,l  \,. \nonumber
\end{align}
It is obvious that ${\bar{a}\ne0}$ when ${\alpha\,l^2\ne0}$. This explicitly demonstrates that the subclass of accelerating (purely) NUT black holes (which have ${a=0}$) is not properly identified in the GP form of the metric \cite{GriffithsPodolsky:2009}, and also in the PV form \cite{PodolskyVratny:2021}. However, for ${\alpha=0}$ we get
${\bar{a}=0}$, ${\bar{l}=\sqrt2\,l}$, and for ${l=0}$ we get ${\bar{a}=0}$, ${\bar{l}=0}$, which are the correct results.

\subsubsection*{$\bullet$ The ${\alpha=0}$ subcase}

\noindent
In this subcase when ${\alpha\to0}$ (so that ${ \tilde{\alpha}=\alpha'= \alpha^2 l^2 \to 0}$), the dominant term in $\kappa$ is
    \begin{align}
        \kappa^2 = \tfrac{1}{4}\alpha^2(l-m+\Lambda l^3)^2 +O(\alpha^4)\,,
    \end{align}
and in $\lambda$ is
    \begin{align}
        \lambda =  \tfrac{1}{4}\alpha^2\big(e^2+g^2-2ml+\tfrac{4}{3}\Lambda l^4 \big) +O(\alpha^4) \,.
    \end{align}
The dominant-order expressions of $\tilde{l}$ and $\tilde{a}$, applying the plus sign in \eqref{l-explicitly} and \eqref{a^2-explicitly}, are thus
    \begin{align}
        \tilde{l}=\dfrac{1}{\alpha l}\,,\qquad
        \tilde{a}=\dfrac{2}{\sqrt{1+\Lambda l^2}}\,.
    \end{align}

Using \eqref{a_0}--\eqref{a_4} for ${a=0}$ and ${\alpha\to 0}$ with
\eqref{direct_transformation_A-PD_parameters-a=0-again} we obtain the coefficients $a_0, a_3, a_4$. There is a miraculous simplification  of these rather complicated expressions to
    \begin{align}
    a_0 = 1+\Lambda l^2 \,, \qquad
    a_3 = 0\,, \qquad
    a_4 =0 \,. \label{a_0-for-a=0-alpha=0}
    \end{align}
The physical (non-dimensionless) parameters of the GP metric are obtained by the rescaling  \eqref{resc} with \eqref{a_0-for-a=0-alpha=0} and $c=\sqrt2\,\alpha\,l^2$ as
\begin{align}
      \bar{\alpha} &= \frac{1}{c}\,\tilde{\alpha} = \frac{\alpha^2 l^2}{\sqrt2\,\alpha\,l^2} = \frac{\alpha}{\sqrt2} \quad\ \to 0 \,,\nonumber\\
      \bar{a} &= \frac{c}{\sqrt{a_0}}\,\tilde{a} = 2\sqrt2\,\,\frac{\alpha\,l^2}{1+\Lambda l^2} \label{a=0-alpha=0}
      \quad\to 0 \,, \\
      \bar{l} &= \frac{c}{\sqrt{a_0}}\,\tilde{l} = \sqrt2\,\,\frac{l}{\sqrt{1+\Lambda l^2}}\,, \nonumber
\end{align}
and
\begin{align}
    &
      \bar{k} = k' = -\frac{2m\,l +e^2+g^2}{4\alpha^2 l^4} - \dfrac{\Lambda}{3\alpha^2}\,, \nonumber \\
    &
      \bar{\epsilon} = \epsilon' = \frac{1}{1+\Lambda l^2} \Big(3\,\frac{e^2+g^2}{2l^2}-2\Big),   \nonumber\\
    &
      \bar{n} = \frac{c}{a_0^{3/2}}\,\,n'
      = \frac{1}{\sqrt2\,(\sqrt{1+\Lambda l^2}\,)^3}\,\Big[
      (m-l)+\frac{e^2+g^2}{l} + \dfrac{\Lambda}{3}\,l^3\Big],\nonumber\\
   &
      \bar{m} = \frac{c}{a_0^{3/2}}\,m'
      = \frac{1}{\sqrt2\,(\sqrt{1+\Lambda l^2}\,)^3}\,\Big[
      -(m+l)+\frac{e^2+g^2}{l} + \dfrac{\Lambda}{3}\,l^3\Big],   \label{resc-a=0,alpha=0}\\
    &
      \bar{e} = \frac{c}{a_0}\,e' = \frac{e}{1+\Lambda l^2} \,, \qquad
      \bar{g} = \frac{c}{a_0}\,g' = \frac{g}{1+\Lambda l^2} \,, \nonumber\\
    &
      \bar{a}_0 = \frac{a_0}{c} = \frac{1+\Lambda l^2}{\sqrt2\,\alpha\,l^2}\,, \qquad
      \bar{a}_3 = \frac{a_3}{a_0} = 0\,, \qquad
      \bar{a}_4 = \frac{a_4}{a_0} = 0 \,.  \nonumber
\end{align}
Notice also that
    \begin{align}
   \bar{\alpha}\,\bar{a}_0 = \dfrac{1+\Lambda l^2}{2l^2}=\dfrac{1}{\,\bar{l}^{\,2}}\,. \label{bar_eg_rel}
    \end{align}
The metric functions \eqref{Omega-final}--\eqref{Q-final} of the GP metric \eqref{GP-metric-final}, namely
\begin{align}
    \dd s^2=
    & -\dfrac{\bar{Q}}{{\bar{\Omega}}^2 \bar{\rho}^{\,2}}
    \big[\, \dd \bar{t} -  2\bar{l}(1-\cos\theta)\dd\bar{\varphi}  \,\big]^2
     +\dfrac{\,\bar{\rho}^{\,2}}{{\bar{\Omega}}^2\bar{Q}}\,\dd\bar{r}^{\,2}
     +\frac{\bar{\rho}^{\,2}}{{\bar{\Omega}}^2} \big(\,\dd\theta^2  + \sin^2\theta\,\dd\bar{\varphi}^2\,\big),
    \label{ds2_NUT-rescaled-alpha=0}
\end{align}
thus become ${\bar{P}=1}$,
\begin{align}
  \bar{\Omega}     &= \dfrac{1}{\bar{l}}\, (\bar{l}-\bar{r}) \,, \qquad
  \bar{\rho}^{\,2} = \bar{r}^{\,2} + \bar{l}^{\,2}\,, \label{Olega and rho_a=0,alpha=0}\\
  \bar{Q} &=  \Big( \bar{e}^{\,2}+\bar{g}^{\,2} + \frac{\alpha^2\bar{k}}{2}\,\bar{l}^{\,4} \Big)
        -2\bar{m}\,\bar{r}+\bar{\epsilon}\,\bar{r}^{\,2}
        -2\,\dfrac{\bar{n}}{\,\bar{l}^{\,2}}\,\bar{r}^{\,3}
        -\Big(\frac{\alpha^2  \bar{k}}{2} + \frac{\Lambda}{3} \Big)\,\bar{r}^{\,4}, \label{Q_a=0,alpha=0}
\end{align}
with the explicit coefficients \eqref{resc-a=0,alpha=0}. This is the exact solution of charged NUT--(anti-)de~Sitter black hole  (without acceleration and Kerr-like rotation). It generalizes the metric given by Eqs.~(147)--(150) in our previous paper \cite{OvcharenkoPodolskyAstorino:2024} to any value of the cosmological constant $\Lambda$ .

To express this solution in the \emph{usual metric form}, it is necessary to perform the transformation of coordinates, and the rescaling of the parameters as
    \begin{align}
        \bar{r} = \dfrac{R-L}{R+L}\,\bar{l}\,,\qquad \bar{t}=\sqrt{2}\,T\,, \qquad
        M=\dfrac{m}{(\sqrt{1+\Lambda l^2}\,)^3} \,, \qquad
        L=\dfrac{\bar{l}}{\sqrt{2}}=\dfrac{l}{\sqrt{1+\Lambda l^2}}\,.\label{LM_rels}
    \end{align}
In such a case,
 \begin{align}
        \dfrac{\bar{\rho}^2}{\bar{\Omega}^2} = R^2+L^2\,,\qquad
        \dfrac{\dd\bar{r}^{\,2}}{\bar{Q}}=\dfrac{\dd R^2}{{\cal Q}}\,,\qquad
        \dfrac{1}{\bar {\Omega}^2}\dfrac{\bar{Q}}{\bar{\rho}^2}=\dfrac{{\cal Q}}{2(R^2+L^2)}\,,
    \end{align}
where
    \begin{align}
        {\cal Q}(R) = \bar{e}^{\,2}+\bar{g}^{\,2} - \dfrac{l^2}{(1+\Lambda l^2)^2} - 2M R
        +\dfrac{1-\Lambda l^2}{1+\Lambda l^2}\,R^2-\dfrac{\Lambda}{3}\,R^4\,.
    \end{align}
Using the relation between  the NUT parameters (an inverse  of \eqref{LM_rels})
    \begin{align}
       l=\dfrac{L}{\sqrt{1-\Lambda L^2}}\,,
    \end{align}
we get ${{\cal Q} = R^2-2MR-L^2+\bar{e}^2+\bar{g}^2-\Lambda \big(\tfrac{1}{3}R^4+2L^2 R^2-L^4\big)}$. The
metric  \eqref{ds2_NUT-rescaled-alpha=0} becomes
\begin{align}
    \dd s^2=
    -F  \big[\,\dd T - 2L(1-\cos\theta)\dd\bar{\varphi} \,\big]^2
     +\dfrac{\dd R^2}{F} + (R^2+L^2)(\,\dd\theta^2  + \sin^2\theta\,\dd\bar{\varphi}^2),
    \label{ds2_NUT-standard}
\end{align}
where
\begin{align}
  F = \frac{R^2-2MR-L^2+\bar{e}^2+\bar{g}^2-\Lambda \big(\tfrac{1}{3}R^4+2L^2 R^2-L^4\big)}{R^2+L^2}\,.\label{F_NUT-standard}
\end{align}
This is the standard metric form of the charged NUT-(anti-)de~Sitter spacetime,  see Eq.~(12.19) in \cite{GriffithsPodolsky:2009}. For ${L=0}$ it simplifies to the Reissner-Nordstr\"om-(anti-)de~Sitter black hole. For ${\Lambda=0}$ it reduces to Eq.~(154) in our previous paper \cite{OvcharenkoPodolskyAstorino:2024}.

\subsubsection*{$\bullet$ The ${l=0}$ subcase}

\noindent
The expansion of $\kappa$ is now
    \begin{align}
        \kappa^2 = \tfrac{1}{4}\alpha^2 m^2 +O(l)\,,
    \end{align}
while $\lambda$ is
    \begin{align}
        \lambda =  \tfrac{1}{4}\alpha^2\big[(1-\alpha^2l^2)(e^2+g^2-2ml)+\tfrac{4}{3}\Lambda l^4 \big]
           \,.
    \end{align}
Using \eqref{direct_transformation_A-PD_parameters-a=0-again}, the dominant-order terms in $\tilde{l}$ and $\tilde{a}$ given by \eqref{l-explicitly} and \eqref{a^2-explicitly} for ${l \to 0}$ give
    \begin{align}
        \tilde{l}=\dfrac{1}{\alpha l}\,,\qquad
        \tilde{a}=2\,.
    \end{align}
A careful evaluation of \eqref{a_0}--\eqref{a_4} in the limit ${l \to  0}$ yields
    \begin{align}
    a_0 = 1 \,, \qquad
    a_3 = 2\alpha m\,, \qquad
    a_4 = -\alpha^2(e^2+g^2) \,. \label{a_0-for-a=0-l=0}
    \end{align}
The physical parameters of the GP metric are obtained by the rescaling  \eqref{resc} of \eqref{direct_transformation_A-PD_parameters-a=0-again} by ${a_0 = 1}$ and $c=\sqrt2\,\alpha\,l^2$ as
\begin{align}
      \bar{\alpha} &= \frac{1}{c}\,\tilde{\alpha} = \frac{\alpha}{\sqrt2} \,,\nonumber\\
      \bar{a} &= c\,\tilde{a} = 2\sqrt2\,\alpha\,l^2 \quad\to 0 \,, \label{limit}\\
      \bar{l} &= c\,\tilde{l} = \sqrt2\,l \qquad \quad\to 0\,, \nonumber
\end{align}
and (in fact, the terms with higher powers of $l$ must be considered in the limit ${l \to 0}$)
\begin{align}
    &
      \bar{k} = k' = \frac{\alpha^2 l^2-1}{4\alpha^2 l^4}\,\big(2m\,l +e^2+g^2   \big) - \dfrac{\Lambda}{3\alpha^2}  \,, \nonumber \\
    &
      \bar{\epsilon} = \epsilon' = 3\,\frac{e^2+g^2}{2l^2} -2 (1+\alpha^2l^2) + \frac{1}{2}\alpha^2(e^2+g^2),   \nonumber\\
    &
      \bar{n} =\, c\,\,n'\,
      = \frac{1}{\sqrt2}\,\Big[ \frac{e^2+g^2}{l} + (m-l) +\alpha^2l^2(m+l) + \dfrac{\Lambda}{3}\,l^3 \,\Big],\nonumber\\
   &
      \bar{m} = c\,m'
      = \frac{1}{\sqrt2}\,\Big[ \frac{e^2+g^2}{l} - (m+l) -\alpha^2l^2(m-l) + \dfrac{\Lambda}{3}\,l^3 \,\Big],   \label{resc-a=0,l=0}\\
    &
      \bar{e} = c\,e' = \sqrt{1-\alpha^2 l^2}\,e \,, \qquad
      \bar{g} = c\,g' = \sqrt{1-\alpha^2 l^2}\,g \,, \nonumber\\
    &
      \bar{a}_0 = \frac{1}{\sqrt2\,\alpha\,l^2}\,, \qquad
      \bar{a}_3 = 2\alpha m\,, \qquad
      \bar{a}_4 = -\alpha^2(e^2+g^2) \,,  \nonumber
\end{align}
so that
    \begin{align}
   \bar{\alpha}\,\bar{a}_0 = \dfrac{1}{2l^2}=\dfrac{1}{\,\bar{l}^{\,2}}\,,\qquad\hbox{and}\quad
   \bar{a} = 2\alpha l\,\bar{l}\,\,. \label{bar_eg_rel2}
    \end{align}

Finally, if we now perform a transformation of coordinates
    \begin{align}
        \bar{r}=\sqrt{2}\dfrac{lR}{R+2l}\,,\qquad
        \bar{\varphi}=\varphi+\alpha\,T\,,\qquad
        \bar{t}=\sqrt{2}\,T\,,
    \end{align}
(see Eq.~(156)  in \cite{OvcharenkoPodolskyAstorino:2024}), the GP metric \eqref{GP-metric-final} in the limit ${l \to 0}$  simplifies to
   \begin{align}
        \dd s^2=\dfrac{1}{(1-\alpha R \cos\theta)^2}\,\Big[-\bar{Q}\,\dd T^2+\dfrac{\dd R^2}{\bar{Q}}+\dfrac{R^2}{\bar{P}}\,\dd\theta^2+\bar{P} R^2 \sin^2 \theta\, \dd\varphi^2\Big],
         \label{ds2_C-metric-standard}
    \end{align}
where
    \begin{align}
        \bar{P}&=1-2\alpha m \cos\theta+\alpha^2(e^2+g^2)\cos^2\theta\,,\\
        \bar{Q}&=(1-\alpha^2 R^2)\Big(1-\dfrac{2m}{R}+\dfrac{e^2+g^2}{R^2}\Big)-\dfrac{\Lambda}{3}R^2\,.
    \end{align}
This is exactly the \emph{standard form of the C-metric solution}, as given in \cite{GriffithsPodolsky:2009} by Eqs.~(14.6), (14.45) for ${a=0}$. For vanishing acceleration (${\alpha=0}$) it reduces to
   \begin{align}
        \dd s^2 &=-\bar{Q}\,\dd T^2+{\bar{Q}}^{-1}\dd R^2+ R^2 (\dd\theta^2+\sin^2 \theta\, \dd\varphi^2) \,,
        \nonumber      \\[2mm]
        & \bar{Q} = 1-\dfrac{2m}{R}+\dfrac{e^2+g^2}{R^2}-\dfrac{\Lambda}{3}R^2\,, \label{RNAdS}
    \end{align}
the usual form of spherically symmetric  Reissner-Nordstr\"om-(anti-)de~Sitter black hole.

\newpage

\section{Conclusions and summary of the subcases}
\label{sc:conclusions}

In this work we investigated a \emph{complete class} of black hole spacetimes of algebraic type type~D characterized by 7~physical parameters, namely their mass $m$, Kerr-like rotation~$a$, NUT parameter~$l$, acceleration~$\alpha$, and electric and magnetic charges $e$ and $g$ (generating electromagnetic field aligned with \emph{both} double-degenerate principal null directions of the Weyl tensor), and any value of cosmological constant $\Lambda$. It is a generalization of results which we previously presented in \cite{OvcharenkoPodolskyAstorino:2024} only for the ${\Lambda=0}$ case.

In particular, we found explicit coordinate transformations and the corresponding relations between the physical parameters of various metric forms of such exact solutions to the Einstein-Maxwell-$\Lambda$ system. In particular, we found the rigorous relations between the (improved) Astorino metric A$^+$, the Pleba\'nski-Demia\'nski metric PD$_{\alpha}$, and the Griffiths-Podolsk\'y metric GP. These are contained in Sections~\ref{sc:PD-form} and~Section~\ref{sc:GP-form}, respectively.

Main conclusions are the following:
\vspace{-1mm}

\begin{itemize}

\item These 3 representations all cover the complete class of such type D black holes.

\item The exact maps between the sets of physical parameters (summarized in Table~\ref{Tab-summary-of-metrics}) are complicated in the most general case. However, there is a considerable simplification in many subcases when (at least) one parameter vanishes.

\item The A$^+$ metric representation is very useful because its parameters can be set to zero, and it can be done in any order. This simple procedure leads to expected special cases, without any unpleasant coordinate degeneracies.

\item The transformation from the A$^+$ form \eqref{ds2_simpl}--\eqref{delta_x_init} to the PD$_{\alpha}$ form \eqref{PD-GP-form}--\eqref{C-definition} is presented in Section~\ref{sc:PD-form}, in particular in Eqs.~\eqref{direct_transformation_A-PD-t}--\eqref{defI-and -defJ} and \eqref{direct_transformation_A-PD_parameters-simplified}--\eqref{coeff_Lambda}. In Subsections~\ref{sc:lambda=0PD}--\ref{sc:a=0PD} we also discussed the special cases ${\Lambda=0}$, ${l=0}$, ${\alpha=0}$, and ${a=0}$.

\item In Section~\ref{sc:GP-form} we gave the subsequent transformation from the PD$_{\alpha}$ form to the GP form    \eqref{GP-metric-final}--\eqref{Q-final}. In Subsections~\ref{sc:lambda=0GP}--\ref{sc:a=0GP} we discussed the special cases ${\Lambda=0}$, ${l=0}$, ${\alpha=0}$, and ${a=0}$ (elusive until recently), and in each of them also the three possible subcases when another physical parameter is zero.

\item We derived explicit and general  expressions for the parameters representing rotation~$\bar{a}$, NUT~$\bar{l}$, and acceleration~$\bar{\alpha}$ in the GP metric form, in terms of the genuine A parameters. These are given by Eqs.~\eqref{a^2-explicitly}, \eqref{l-explicitly}, and \eqref{direct_transformation_A-PD-acceleration-explicit-again} --- with the rescaling \eqref{resc}.

\item We proved that in many subcases these physical parameters are the same in the GP and A metrics \eqref{GP-metric-final} and \eqref{ds2_simpl}, that is ${\bar{\alpha}=\alpha}$, ${\bar{a}=a}$, ${\bar{l}=l}$. But in some subcases they differ, in particular when ${\Lambda a^2\ne0}$ and ${\Lambda l^2\ne0}$. Moreover, in the subcase of accelerating (purely) NUT black holes (which have ${a=0}$) we get ${\bar{a}\ne0}$ when ${\alpha\,l^2\ne0}$. Such relations are contained in Table~\ref{Tab-summary-of-parameters}.
\end{itemize}
\vspace{-2mm}

\renewcommand{\arraystretch}{1.0}
\begin{table}[!h]
\begin{center}

\begin{tabular}{|c||c|c|c|c|}
\hline
             & \ case \ & \ case \     &  \ case \ \\[-8pt]
\ subcase\ \ & ${l=0}$  & ${\alpha=0}$ &  ${a=0}$  \\[2pt]
\hline
\hline
&
& ${\bar{\alpha}=0}$
& ${\bar{\alpha}_s = \alpha}$  \\
${l=0}$
& ---
& ${\bar{l}=0}$
& ${\bar{l}_s=0}$  \\
&
& ${\,\bar{a}=\dfrac{\sqrt{1+\tfrac{4}{3}\Lambda a^2}-1}{\tfrac{2}{3}\Lambda a}\,}$
& ${\bar{a}_s=0}$  \\
\hline
&
${\bar{\alpha}=0}$
&
& ${\bar{\alpha}_s=0}$  \\
${\alpha=0}$
& ${\bar{l}=0}$
& ---
& ${\,\bar{l}_s=\dfrac{l}{\sqrt{1+\Lambda l^2}}\,}$  \\
& ${\,\bar{a}=\dfrac{\sqrt{1+\tfrac{4}{3}\Lambda a^2}-1}{\tfrac{2}{3}\Lambda a}\,}$
&
& ${\bar{a}_s=0}$  \\
\hline
& ${\bar{\alpha}=\alpha}$
& ${\bar{\alpha}=0}$
&  \\
${a=0}$
& ${\bar{l}=0}$
& ${\bar{l}=\dfrac{l}{\sqrt{1+\Lambda l^2}}}$
& --- \\
& ${\bar{a}=0}$
& ${\bar{a}=0}$
& \\
\hline
& ${\bar{\alpha}=\alpha}$
& ${\bar{\alpha}=0}$
& ${\bar{\alpha}_s = \alpha}$   \\
${\Lambda=0}$
& ${\bar{l}=0}$
& ${\bar{l}=l}$
& ${\,\bar{l}_s=\dfrac{1}{\sqrt{a_0}}\,\dfrac{1+\alpha^2 l^2}{1-\alpha^2 l^2}\,l\,}$  \\
& ${\bar{a}=a}$
& ${\bar{a}=a}$
& ${\,\bar{a}_s=\dfrac{1}{\sqrt{a_0}}\,\dfrac{2\,\alpha\,l^2}{1-\alpha^2 l^2}\,}$  \\
\hline
\end{tabular}
\vspace{4.0mm}
\caption{\label{Tab-summary-of-parameters} Summary of the relations between the A and GP physical parameters in the most important subcases. The constant $a_0$ in the last column of the last row is given by Eq.~\eqref{a_0-for-a=0-Lambda=0}.
In the last column corresponding to the case ${a=0}$ we use the trivially re-scaled physical parameters
${\,\bar{\alpha}_s := \sqrt{2}\,\bar{\alpha}}$, ${\,\bar{l}_s := \bar{l}/\sqrt{2}}$, ${\,\bar{a}_s := \bar{a}/\sqrt{2}}$. Actually, ${\bar{l}_s = L}$ introduced in the transformation \eqref{LM_rels}. Clearly, all the subcases commute with the cases (i.e., the results do not depend on the order in which the two limits are performed).}
\end{center}
\end{table}

\newpage

\section*{Acknowledgments}

This work was supported by the Czech Science Foundation Grant No.~GA\v{C}R 23-05914S (HO, MA) and by the Czech Science Foundation Grant No.~GA\v{C}R 22-14791S (JP).

\end{document}